\documentclass[structabstract]{aa}
\pdfoutput=1
\usepackage{subfig}
\usepackage{graphicx}
\usepackage{txfonts}
\usepackage{natbib}
\usepackage{url}
\usepackage{lscape}
\usepackage{booktabs}

\newcommand{\mJybeam}{mJy beam$^{-1}$}

\newcommand{\kms}{$\,$km$\,$s$^{-1}$}
\newcommand{\kpc}{kpc}

\bibpunct{(}{)}{;}{a}{}{,}

\begin{document}

\title{AGN duty cycle estimates for the ultra-steep spectrum radio relic VLSS~J1431.8+1331}

\titlerunning{LOFAR studies of VLSS~J1431.8+1331}
\authorrunning{Shulevski et al.}	
\author{A.~Shulevski\inst{1,2}\and
	R.~Morganti\inst{1,2}\and 
	P.~D.~Barthel\inst{1}\and
	J.~J.~Harwood\inst{2}\and
	G.~Brunetti\inst{3}\and
	R.~J.~van~Weeren\inst{4}\and
	H.~J.~A.~R\"{o}ttgering\inst{5}\and
	G.~J.~White\inst{9,10}\and
	C.~Horellou\inst{8}\and
	M.~Kunert-Bajraszewska\inst{11}\and
	M.~Jamrozy\inst{12}\and
	K.~T.~Chyzy\inst{12}\and
	E.~Mahony\inst{2}\and
	G.~Miley\inst{5}\and
	M.~Brienza\inst{2,1}\and
	L.~B\^{\i}rzan\inst{6}\and
	D.~A.~Rafferty\inst{6}\and
	M.~Br\"{u}ggen\inst{6}\and
	M.~W.~Wise\inst{2,7}\and
	J.~Conway\inst{8}\and
	F.~de~Gasperin\inst{6}\and
	N.~Vilchez\inst{2}
	}
	\institute{University of Groningen, Kapteyn Astronomical Institute, Landleven 12, 9747 AD Groningen, The Netherlands\\
		   \email{shulevski@astron.nl, pdb@astro.rug.nl}\and
		   ASTRON, the Netherlands Institute for Radio Astronomy, Postbus 2, 7990 AA, Dwingeloo, The Netherlands\\
		   \email{morganti@astron.nl}\and
		   IRA-INAF, via P. Gobetti 101, 40129 Bologna, Italy\and
		   Harvard-Smithsonian Center for Astrophysics, 60 Garden Street, Cambridge, MA 02138, USA\and
		   Leiden Observatory, Leiden University, Niels Bohrweg 2, 2333 CA Leiden, The Netherlands\and
		   Universit\"{a}t Hamburg, Hamburger Sternwarte, Gojenbergsweg 112, D-21029, Hamburg, Germany\and
		   Astronomical Institute ’Anton Pannekoek’, University of Amsterdam, Postbus 94249, 1090 GE Amsterdam, The Netherlands \and
		   Department of Earth and Space Sciences, Chalmers University of Technology, Onsala Space Observatory, 43992, Onsala, Sweden\and
		   Department of Physics and Astronomy, The Open University, Milton Keynes MK7 6AA, England\and
		   RAL Space, The Rutherford Appleton Laboratory, Chilton, Didcot, Oxfordshire OX11 0QX, England\and
		   Toru\'n Centre for Astronomy, Faculty of Physics, Astronomy and Informatics, NCU, Grudziacka 5, 87-100 Toru\'n, Poland\and
		   Obserwatorium Astronomiczne, Uniwersytet Jagiello\'{n}ski, ul Orla 171, 30-244, Krak\'{o}w, Poland
		  }
\date{\today}
	\abstract
	{Steep spectrum radio sources associated with active galactic nuclei (AGN) may contain remnants of past AGN activity episodes. Studying these sources gives us insight into the AGN activity history. Novel instruments like the LOw Frequency ARray (LOFAR) are enabling studies of these fascinating structures to be made at tens to hundreds of MHz with sufficient resolution to analyse their complex morphology.}
	{Our goal is to characterize the integrated and resolved spectral properties of VLSS~J1431+1331 and estimate source ages based on synchrotron radio emission models, thus putting constraints on the AGN duty cycle.}
	{Using a broad spectral coverage, we have derived spectral and curvature maps, and used synchrotron ageing models to determine the time elapsed from the last time the source plasma was energized. We used LOFAR, Giant Metrewave Radio Telescope (GMRT) and Jansky Very Large Array (VLA) data.}
	{We confirm the morphology and the spectral index values found in previous studies of this object. Based on our ageing analysis, we infer that the AGN that created this source currently has very low levels of activity or that it is switched off. The derived ages for the larger source component range from around 60 to 130 Myr, hinting that the AGN activity decreased or stopped around 60 Myr ago. We observe that the area around the faint radio core located in the larger source component is the youngest, while the overall age of the smaller source component shows it to be the oldest part of the source.}
	{Our analysis suggests that VLSS~J1431.8+1331 is an intriguing, two-component source. The larger component seems to host a faint radio core, suggesting that the source may be an AGN radio relic. The spectral index we observe from the smaller component is distinctly flatter at lower frequencies than the spectral index of the larger component, suggesting the possibility that the smaller  component may be a shocked plasma bubble. From the integrated source spectrum, we deduce that its shape and slope can be used as tracers of the activity history of this type of steep spectrum radio source. We discuss the implications this conclusion has for future studies of radio sources having similar characteristics.}

\maketitle

\section{Introduction}
\label{c4:intro}

Active galactic nuclei (AGN) are cosmic powerhouses that produce prodigious amounts of energy and deposit it into the interstellar and the intergalactic medium (ISM, IGM). To estimate the total energy output of an AGN over cosmic time, it is necessary to determine its duty cycle, defined as the ratio of the time intervals during which the AGN is active and shut down. This is important, since we have a limited insight into the AGN duty cycle \citep[][]{RefWorks:34, RefWorks:178}, while we have some knowledge of the energy output of an AGN during the duration of the active phase \citep[][]{RefWorks:42, RefWorks:43}. Evidence of multiple episodes in radio AGN activity has been growing steadily \citep[][]{RefWorks:28, RefWorks:262, RefWorks:6, RefWorks:2}. \cite{RefWorks:96} have shown that the period of time that an AGN is active is $ 10^{7} - 10^{8} $ years in their source sample of several AGN radio relics, while the time between the active phases is an order of magnitude shorter. The missing pieces of information are important for various reasons. The (total) energy budget is important for studying galactic evolution because the AGN energy output influences the cooling of gas and galactic assembly. Another important area of astrophysics that is affected is the star formation history in the host galaxy, since AGN "feedback" can potentially quench star formation through influencing the ISM. The AGN fuelling \citep[][]{RefWorks:105, RefWorks:143} can potentially also be interrupted.

An observational signature of a "switched off" AGN is the steep spectrum of the associated radio emission at high frequencies ($ \alpha < -1.5 $)\footnote{We use the $ S \propto \nu ^{\alpha} $ definition for the spectral index $ \alpha $ throughout this work.}, leading to brighter emission at low frequencies. The reason for this is that to first order (neglecting other energy loss mechanisms), the plasma ejected from the AGN during its active phase will mainly lose energy through synchrotron radiation and inverse Compton (IC) scattering of electrons off cosmic microwave background (CMB) photons. As a consequence, high energy particles that radiate the most at high radio frequencies lose their energy fastest. This produces a characteristic synchrotron radio spectrum that evolves over time in such a way that most of the particles that still radiate after the AGN has shut down are low energy particles, and the radio emission is strongest at low frequencies. Thus, low frequency radio surveys are the best tool for detecting such sources.

The organization of the paper is as follows. In section \ref{c4:intro:target} we describe our target in some detail and outline the goals of our study. Section \ref{c4:obsdata} describes our data set and gives an overview of the data reduction procedures. In Section \ref{c4:res} we present the results of the data analysis, and we discuss our findings in Section \ref{c4:disc}. We finish by presenting our conclusions in Section \ref{c4:fin}.

\section{VLSS~J1431.8+1331}
\label{c4:intro:target}

\begin{figure}[!htpb]
  \centering
  \includegraphics[width=90mm]{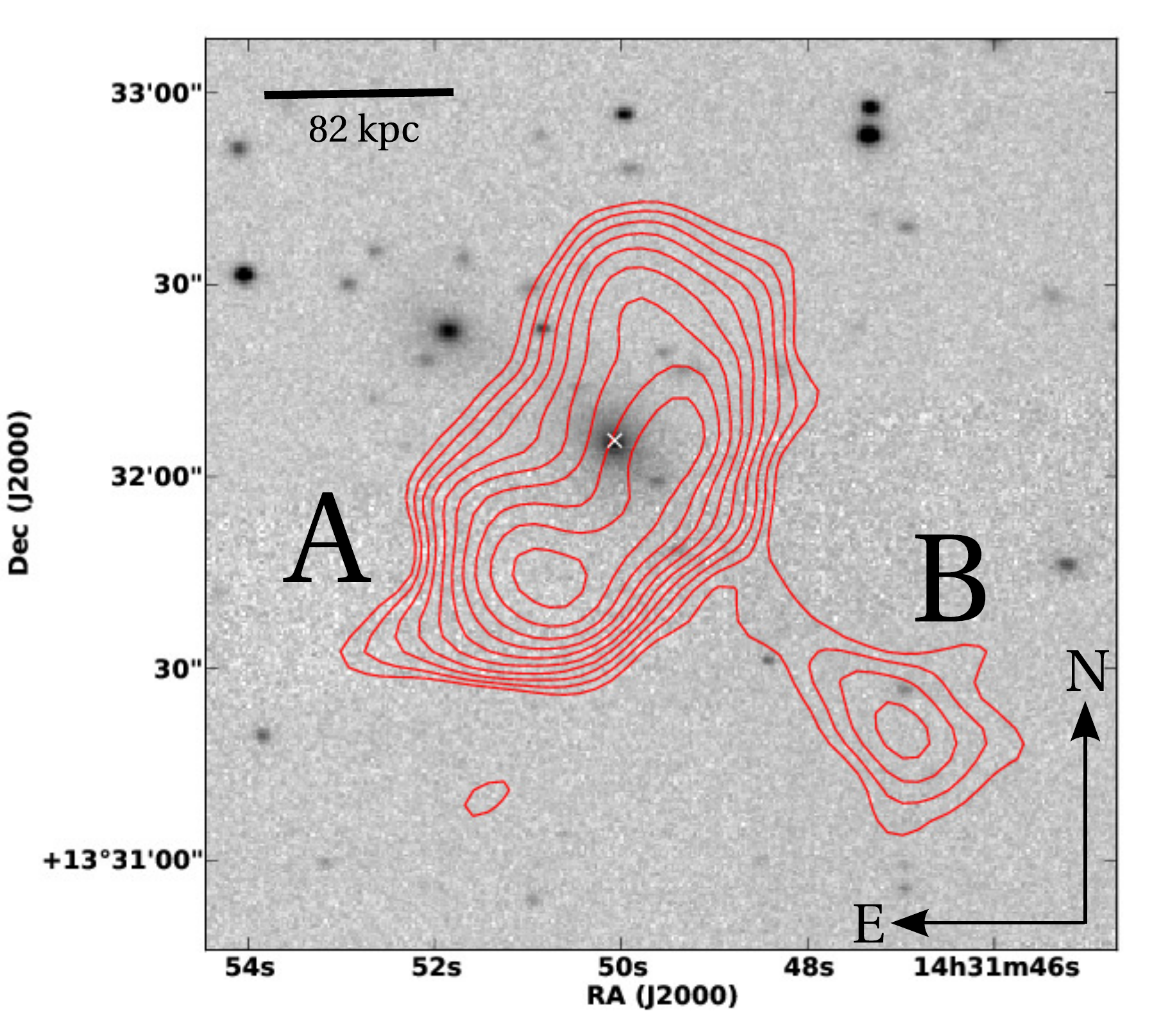}
  \caption{LOFAR HBA image contours (red) of VLSS~J1431.8+1331 (centred on 148 MHz and with a bandwidth of 48 MHz) superposed on a Sloan Digital Sky Survey (SDSS) greyscale image of the galaxy cluster maxBCG~J217.95869+13.53470. Eleven log-spaced contour levels are shown in red, spanning the surface brightness interval between $ -10 $ and $ 100 $ \mJybeam, with a beam size of $ 10\farcs5 \times 8 \arcsec $. The white cross marks the position of the bright cluster galaxy (BCG) AGN host, and the source components are labelled.}
\label{c4:SDSS_LOFAR}
\end{figure}

Several steep spectrum sources were studied by \cite{RefWorks:54, RefWorks:145}, identified by matching the VLSS and NVSS\footnote{VLSS is the VLA Low frequency Sky Survey carried out at 74 MHz \citep{RefWorks:128}. NVSS stands for the NRAO VLA Sky Survey carried out at a frequency of 1.4 GHz \citep{RefWorks:139}} catalogues. One of these, VLSS J1431.8+1331, was identified as being potentially an AGN radio relic. These structures are radio remnants of a past AGN activity episode, not to be confused with radio halos and relics, which are a consequence of particle acceleration and ensuing radio synchrotron emission during galaxy cluster mergers. The object has an ultra steep ($ \alpha \simeq -2 $) spectrum at high frequencies with a radio morphology pointing towards a complex activity history. It presents two distinct source regions, a larger one and a smaller one (henceforth labelled A and B, respectively), connected by a faint bridge of radio emission (Figure \ref{c4:SDSS_LOFAR}). Region A appears to contain a faint radio core and has a steep spectrum showing little spectral curvature compared to Region B whose spectrum flattens out at low frequencies.

The host galaxy of VLSS~J1431.8+1331 is the brightest cluster galaxy (BCG) of the galaxy cluster maxBCG~J217.95869+13.53470 \citep{RefWorks:144}, located at a redshift\footnote{The adopted cosmology in this work is $ H_{0} = {70.5} $ \kms Mpc$ ^{-1} $, $ \Omega_{M} =  0.27 $, $ \Omega_{\Lambda}  =  0.73 $. At the redshift of VLSS~J1431+1331.8, $ 1\arcsec \, = \, 2.747 $  kpc \citep{RefWorks:155}.} of $ z = 0.1599 $.
The cluster has been observed in X-rays by \cite{RefWorks:86} using XMM Newton. They derived temperature excesses to the north and west of the BCG and a slight increase in entropy coinciding with the south-west component of radio emission. Based on these observations, they propose that maxBCG~J217.95869+13.53470 has interacted with a smaller group of galaxies located to the north-east of the BCG. According to the analysis presented, the smaller galaxy group came from the south-east, and we are observing the system a short while after the closest approach. Shocks related to this process might have compressed a bubble of plasma expelled by the AGN and revived its synchrotron emission, thus giving rise to the smaller region of radio emission to the south-west.

In this work, we aim to investigate whether Region B represents a merger shock rejuvenated synchrotron emission region. We also want to derive more information about the physical nature of the various source components of VLSS~J1431.8+1331. The co-existence of a (possible) rejuvenated plasma bubble and an AGN relic provides us with a serendipitous chance to compare their spectral properties (at high resolution and low frequencies). Using observations made with the LOFAR telescope's \citep[][]{RefWorks:157} high band antennas, we perform spectral index mapping down to around $ 140 $ MHz with a resolution comparable to that of the Giant Metrewave Radio Telescope (GMRT) and the Jansky Very Large Array (VLA) study made by \cite{RefWorks:54} at higher frequencies. The resulting broad-band data set has enabled us to study the ageing of the plasma in detail across the source. By doing this, we can put constraints on the timescales involved in the past activity of the AGN responsible for the radio emission, and compare our findings with the results of previous studies.

\section{Observations and data reduction}
\label{c4:obsdata}

The target was observed on the night of February 17, 2013 for a total on source time with LOFAR's high band antennas (HBA) of 5.7 hours. The HBA observation was taken in interleaved mode, i.e. 3C~295 was observed as a calibrator source for two minutes, followed by a scan of the target of 11-minute duration with a one-minute gap between calibrator and target scans, allowing for beam forming and target re-acquisition. The 325 sub-bands were recorded covering 64 MHz of bandwidth between 116 MHz and 180 MHz. Each sub-band has 64 frequency channels spanning a bandwidth of 200 kHz. The integration time was set to two seconds for both calibrator and target. Four polarizations were recorded. The HBA station field of view (FoV) spans around four degrees full width at half maximum (FWHM).

\begin{table}[!htpb]
\noindent \caption{\small LOFAR observation configuration}
\label{c4:table:1}
\small
\begin{tabular}{ p{4cm} p{4cm}}
\hline\hline
Central frequency [MHz] & 140 \\
Bandwidth [MHz] & 64 \\
Integration time & 2 seconds\\
Observation duration & 6 hours\\
Polarization & Full Stokes \\
UV range & $ 0.25 k\lambda \, - \, 20 k\lambda $\\
\hline
\end{tabular}
\end{table}

The data were pre-processed by the observatory pipeline \citep{RefWorks:180} as follows. Each sub-band was automatically flagged for RFI using the AOFlagger \citep{RefWorks:133} and averaged in time to ten seconds per sample and in frequency by a factor of 16, which gives us four channels per sub-band in the output data. The calibrator data were used to derive amplitude solutions for each station using the BlackBoard Self calibration - BBS \citep{RefWorks:182} tool that takes the time and frequency varying LOFAR station beams into account. The calibrator flux density scale was set according to \cite{RefWorks:181}. The obtained (complex) station gain solutions were used to fix the target amplitude scale.

Only the Dutch LOFAR stations were used. The amplitude-corrected target visibilities were phase-(self)calibrated incrementally, using progressively longer baselines to get to the final resolution (Vilchez et al. in prep.). The initial phase calibration model was derived from the VLSS catalogue covering the FoV out to the first null of the station beam. It included spectral index information for each source. Before initializing the calibration, we have concatenated the data into 4 MHz (20 SB) groups previously averaging each sub-band to one frequency channel. We have chosen this setup to maximize the S/N while maintaining frequency dependent ionospheric phase rotation to a manageable level. We did not perform any directional solving and did not explicitly correct for ionospheric (direction dependent) effects.

The station beams are complex-valued and time-, frequency-, and direction-dependent, and are not the same for all of the stations. The imaging was done by using the LOFAR imager \citep{RefWorks:183}, which incorporates the LOFAR beam and uses the A-projection \citep{RefWorks:184} algorithm to image the entire FoV. The imager does not (at this stage) implement spectral index correction when it does the multi-frequency imaging. We used Briggs weights \citep{RefWorks:185} with the robustness parameter set to $ -2 $ (uniform weights) and performed a UV plane selection to include all of the baselines to achieve the highest possible resolution. Discarding 8 MHz from either edge of the band, we have imaged the data so as to obtain a final HBA data set of six images of 8 MHz bandwidth each. We	 averaged these images together to obtain one broadband image. Deconvolution artefacts are present around the sources in the image, mostly due to the residual calibration errors caused by the ionosphere. The observation was performed at night when the ionospheric conditions are expected to be most favourable. Based on our inspection of the phase calibration solutions for the target field, we judged the ionospheric conditions to be predominantly calm for the duration of the observing run.

Given the low declination of the target, some uncertainties still exist about the accuracy of the LOFAR in-band spectral index. Thus, we have decided to use only two LOFAR HBA images (given in table \ref{c4:table:2}) in our spectral analysis.

In addition to our LOFAR data, we use GMRT and VLA data kindly supplied by RvW to constrain the spectral index and curvature. The properties of the dataset are listed in Table \ref{c4:table:2}. For the purposes of our work, we list the properties of the smoothed GMRT and VLA images (adapted to the LOFAR HBA synthesized beam size). The r.m.s. noise values for each of the flux measurements were derived according to \cite{RefWorks:141}, taking the background noise away from bright sources and scaling for the contribution of the size of the measurement region used on the target into consideration. To account for the flux uncertainties stemming from the 3C~295 flux scale and the imperfections of the beam model, we added 20\% of the measured flux density value in quadrature to the derived noise in the case of the LOFAR measurements. The same was done for the GMRT measurements, but the flux density value added was 5\% \citep{RefWorks:186}. The details of the GMRT and VLA data sets and the reduction procedure can be found in \cite{RefWorks:54, RefWorks:145}.

\begin{table}[!htpb]
\noindent \caption{\small Details about the images used in our analysis.}
\label{c4:table:2}
\small
\begin{tabular}{p{1.5cm} p{1.5cm} p{1.5cm} p{2.5cm}}
\hline\hline
\small Instrument & \small $ \nu $ [MHz] & \small $ \Delta \nu $ [MHz] & \small $ \sigma $ [\mJybeam] \\
\hline
LOFAR\tablefootmark{a} & 144 & 48 & 0.5 \\
LOFAR\tablefootmark{a} & 135 & 8 & 2.3 \\
LOFAR\tablefootmark{a} & 145 & 8 & 1.8 \\
GMRT\tablefootmark{b}\tablefootmark{d} & 325 & 32 & 0.1 \\
GMRT\tablefootmark{b}\tablefootmark{d} & 610 & 32 & 0.1 \\
VLA\tablefootmark{b}\tablefootmark{d} & 1425 & 100 & 0.05 \\
VLA\tablefootmark{c}\tablefootmark{d} & 1425 & 100 & 0.02 \\
\hline
\end{tabular}
\tablefoot{
\tablefoottext{a}{Beam size: $ 10\farcs5 \times 8\arcsec $.}
\tablefoottext{b}{Smoothed to the LOFAR beam size.}
\tablefoottext{c}{Original VLA resolution of $ 1\farcs6 \times 1\farcs5 $.}
\tablefoottext{d}{Details about the data set and imaging setup can be found in \cite{RefWorks:145}.}
}
\end{table}

\section{Results}
\label{c4:res}

\subsection{Radio morphology}
\label{c4:morph}

\begin{figure*}[!htpb]
\centering
\begin{minipage}[c]{0.5\linewidth}
\centering \includegraphics[width=\textwidth]{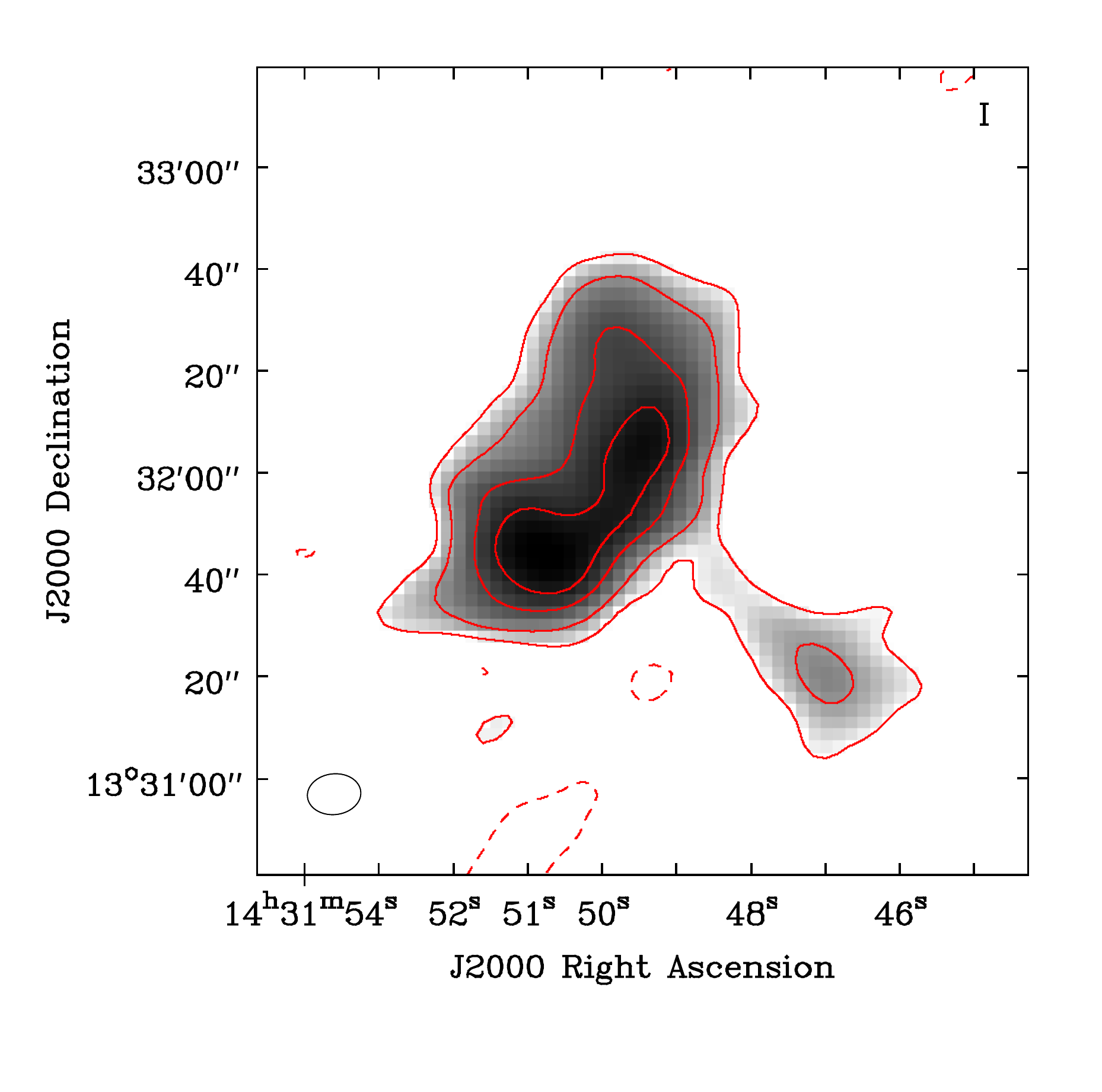}
\end{minipage}%
\begin{minipage}[c]{0.53\linewidth}
\centering \includegraphics[width=\textwidth]{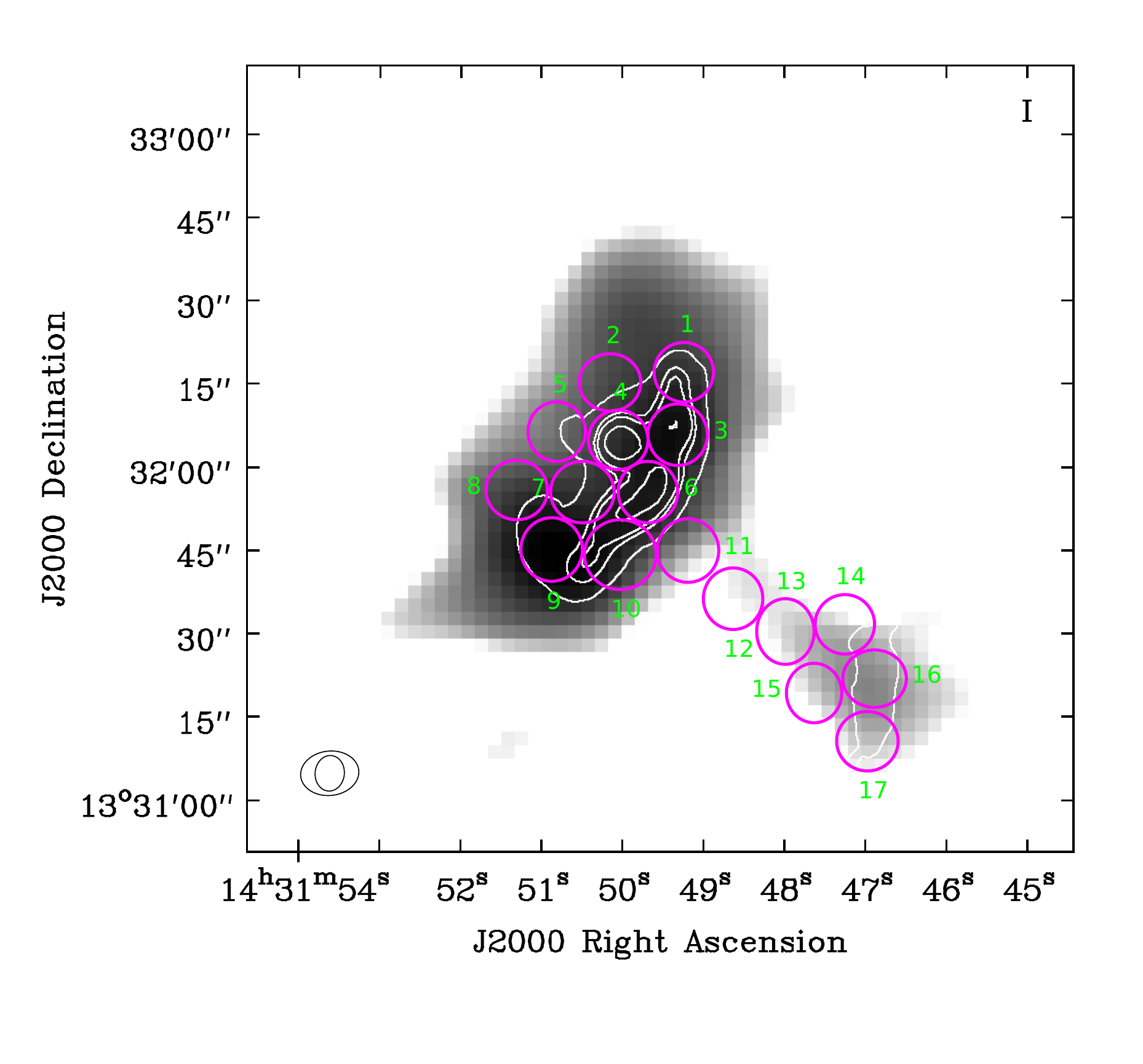}
\end{minipage}
\caption{\textbf{Left:} LOFAR HBA greyscale image of VLSS J1431.8+1331. Overlaid contours (in red): $ [-6, 20, 40, 90, 150] \times \sigma $, $ \sigma = 0.5 $ \mJybeam . \textbf{Right:} Measurement regions (outlined in magenta) used in the spectral (ageing) analysis. VLA 1425 MHz high resolution contours \citep{RefWorks:145} are shown in white, outlining a core inside Region 4. Contour levels: $ [-3, 3, 9, 12, 20] \times 4\cdot10^{-2} $ \mJybeam . The beam sizes are indicated in the lower left corner.}
\label{c4:tgt_regions}
\end{figure*}

Figure \ref{c4:tgt_regions} shows the LOFAR image of the target spanning the HBA band. We can discern two different source regions \citep[described in detail by][]{RefWorks:145}, the larger and brighter one (A) to the north-east, and a smaller region (B) to the south-west connected by a faint "bridge" of radio emission. Both regions seem to be slightly curved to the north-east. This is more noticeable for Region B in the higher resolution image, outlined in contours in the right-hand panel. A hint of a faint radio core is visible in the high resolution VLA image of \cite{RefWorks:145}, while the brightest part of the target in the LOFAR image is the south-eastern part of Region A.

The right-hand panel in Figure \ref{c4:tgt_regions} shows the flux density measurement regions used in the subsequent analysis. The measurement areas have been chosen to be of similar size to the synthesized LOFAR beam (around 24.5 kpc across). We have measured the flux density in each area using our maps (LOFAR, GMRT, and VLA), thus covering about a factor of 10 in frequency range (144 to 1425 MHz). All of the maps were convolved to the same restoring beam size as that of the HBA beam of $ 10\farcs5 \times 8\arcsec $.

\subsection{Spectral analysis and radiative ages}
\label{c4:spec}

In what follows, we use the LOFAR measurements to extend the spectral index analysis to the lowest frequencies ever for this object. Our goal is to trace the lowest energy particle population to better constrain the radio spectra at low frequencies. We also map the radiative ages over the source surface by fitting synchrotron ageing models to our data.

\subsubsection{Spectral index and curvature}
\label{c4:specidxcrv}

\begin{figure*}[!htpb]
\centering
\subfloat[][$ \alpha_{144}^{325} $ spectral index map of VLSS J1431+1331.]{\includegraphics[width=0.5\textwidth]{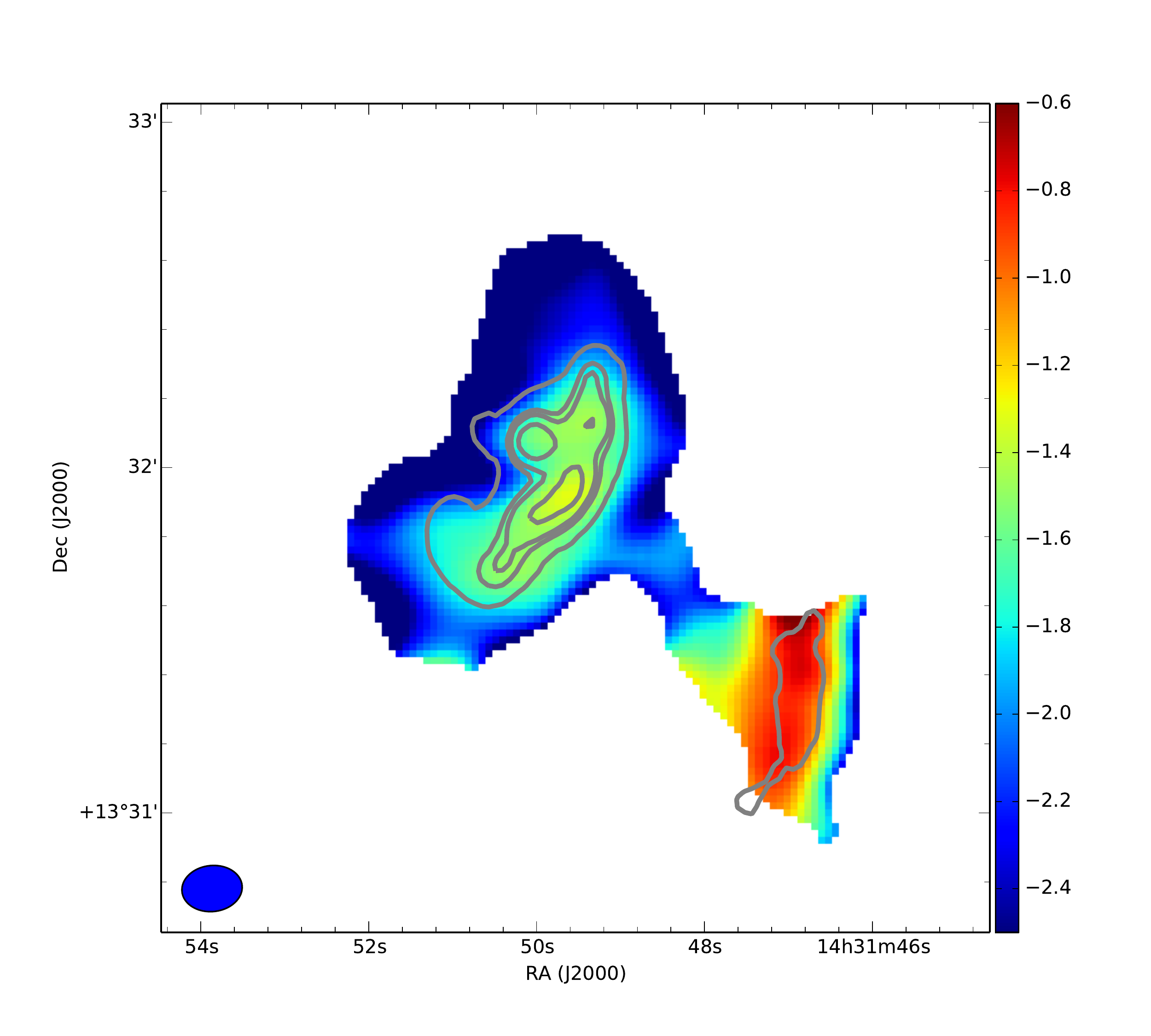}
\label{c4:spix_curv_maps:subfig1}}
\subfloat[][Spectral index error]{\includegraphics[width=0.48\textwidth]{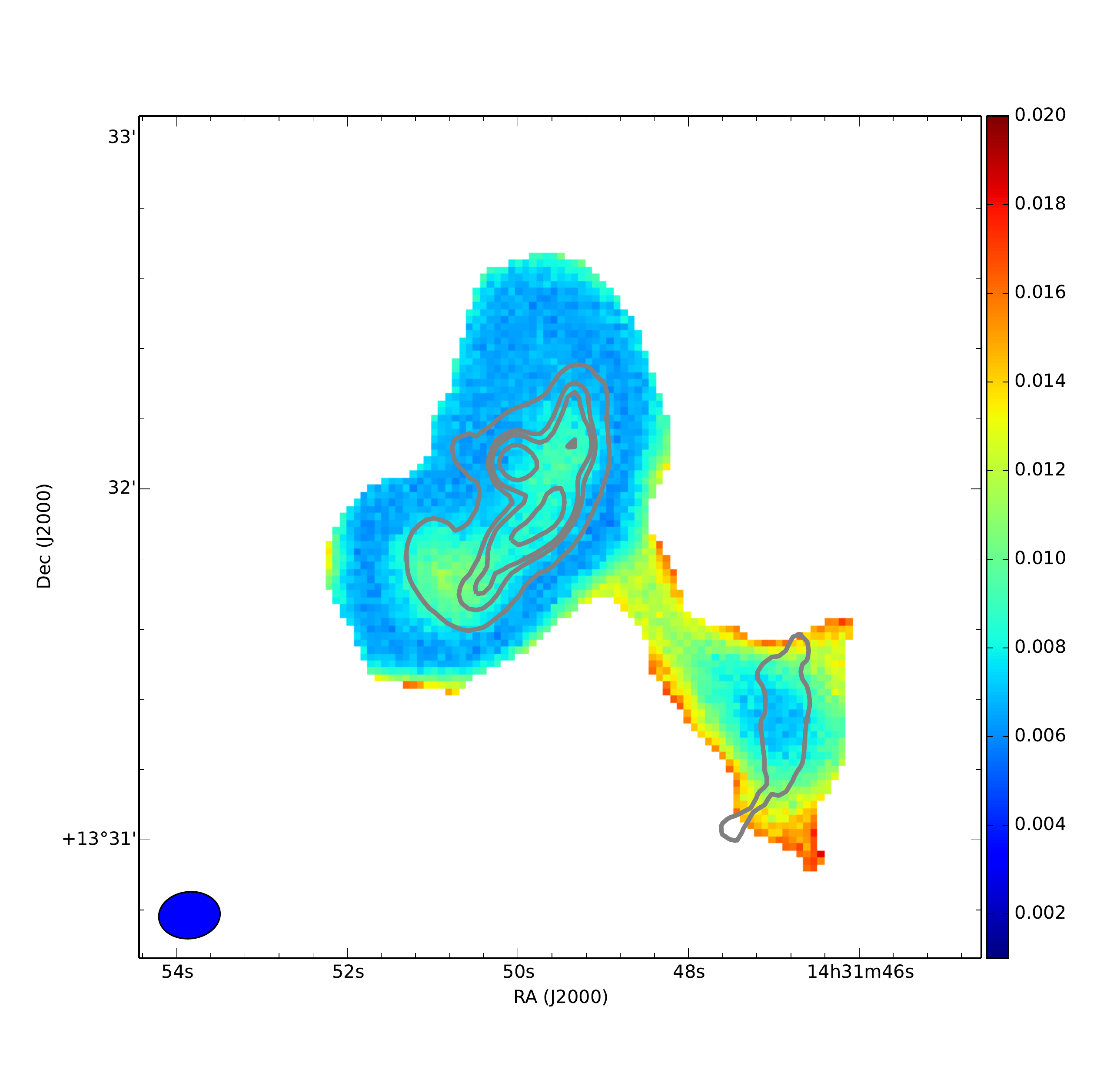}
\label{c4:spix_curv_maps:subfig2}}
\quad
\subfloat[][$ \alpha_{144}^{610} \, - \, \alpha_{610}^{1425} $ spectral curvature map using LOFAR, GMRT and VLA images.]
{\includegraphics[width=0.5\textwidth]{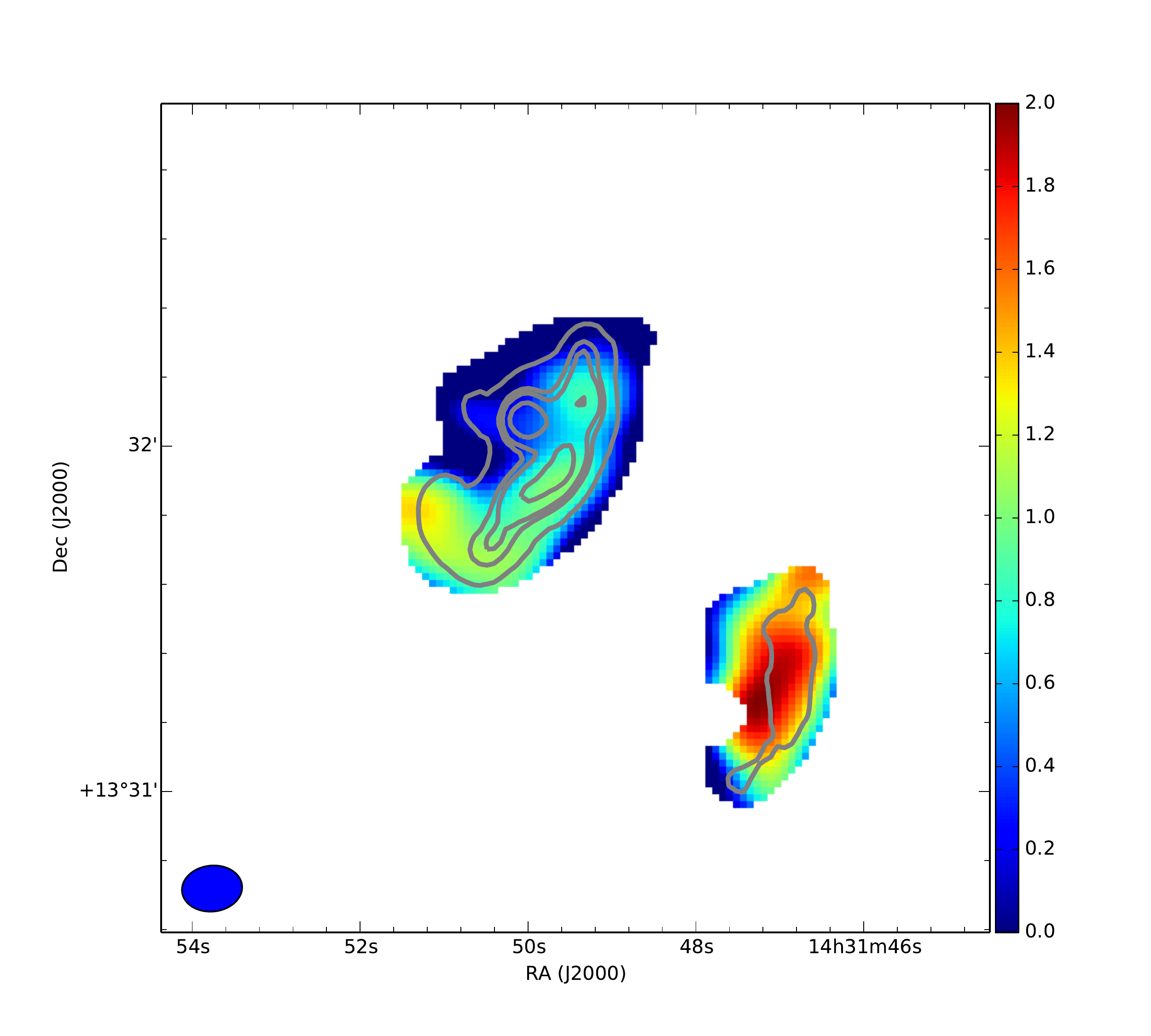}
\label{c4:spix_curv_maps:subfig3}}
\subfloat[][Spectral curvature error]{\includegraphics[width=0.5\textwidth]{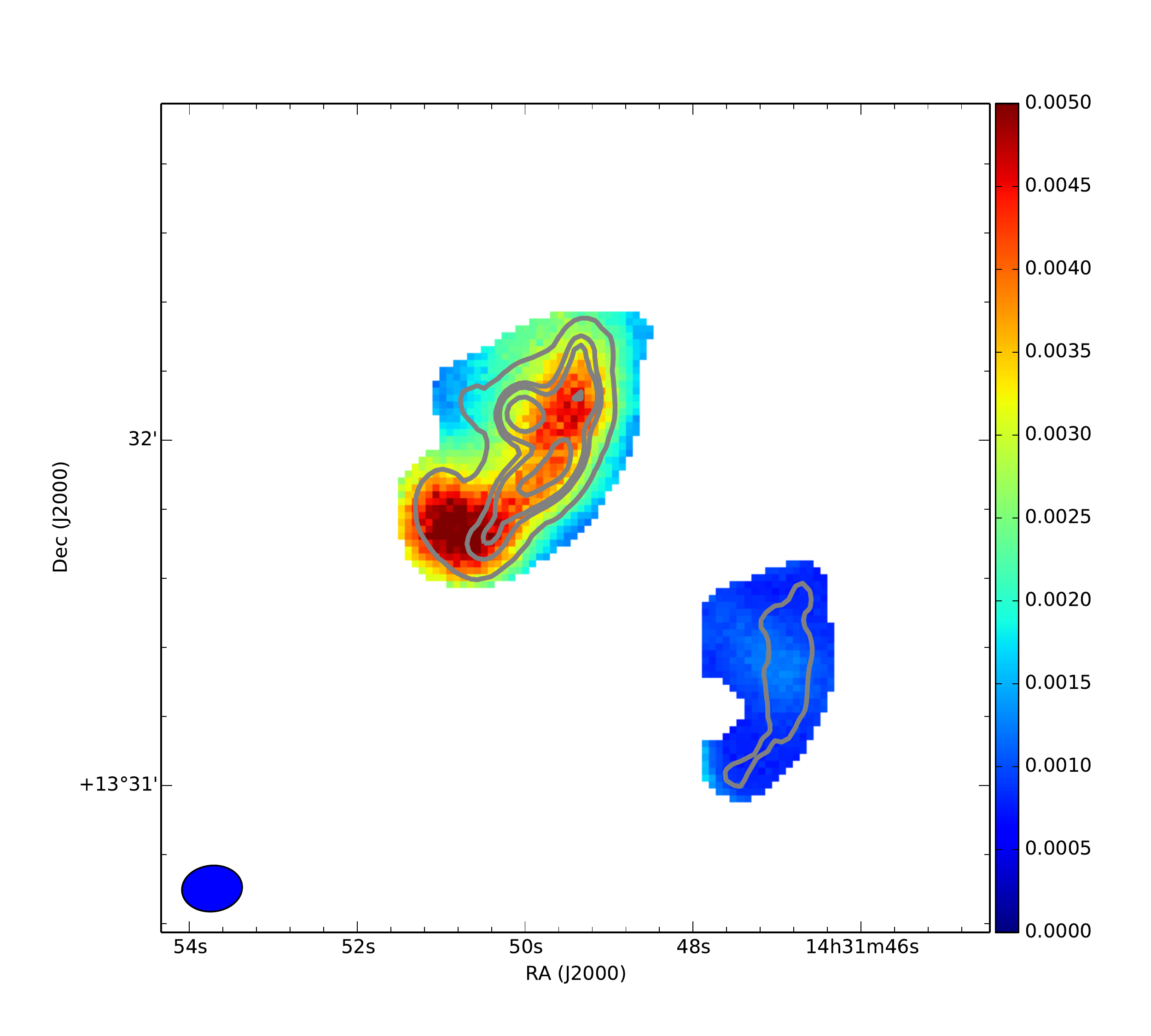}
\label{c4:spix_curv_maps:subfig4}}
\caption{In all panels $ 1425 \, MHz $ high resolution VLA contours overlaid (levels as in Figure \ref{c4:tgt_regions}). The beam size is indicated in the lower left corner.}
\label{c4:spix_curv_maps}
\end{figure*}

We have derived spectral index and spectral curvature ($ \mathrm{SPC} = \alpha_{\mathrm{low}} - \alpha_{\mathrm{high}} $) maps of the target. In Figure \ref{c4:spix_curv_maps} we present the spectral index derived from the LOFAR HBA, GMRT, and VLA maps and the corresponding spectral curvature across the source. The spectral indices were derived by modified\footnote{We drew 100 samples from the interval around the flux density points bounded by the error bars assuming uniform probability distribution. We then fitted for the spectral index; the mean and standard deviation of the fits for the sample are shown in the corresponding maps.} least squares fitting of a power law to the data for each pixel of the maps. Pixels were blanked that had a value below $ 15 \sigma $ for each map when deriving the spectral index and $ 3 \sigma $ when deriving the spectral curvature. This was mainly done to avoid the artefacts present around the target in the LOFAR HBA images. The maximum UV distance of the LOFAR data of 20 $ k\lambda $ corresponds to the one of the GMRT and VLA of 22 $ k\lambda $. The minimum baseline length of the used data sets was 0.25 $ k\lambda $ (the VLA data were taken in the A + B + C array configuration); i.e., the largest spatial scale that can be detected in the images is 13.7 \arcmin.

Region A shows a relatively steep low frequency spectral index of $ \alpha \, \sim \, -1.2 $ around the radio core and the western edge. The spectral index steepens to around $ \alpha \, \sim \, -2 $ going towards the edges. The spectral curvature map indicates that for this source component, the spectrum is flat around the radio core, with breaks at higher frequencies developing at the western edge and in the south-east.

\subsubsection{Ages}
\label{c4:ages}

Synchrotron spectral ageing theory was established by \cite{RefWorks:126}. Later, these foundations were expanded by \cite{RefWorks:187, RefWorks:125}, \cite{RefWorks:278, RefWorks:188, RefWorks:82}, and \cite{RefWorks:34} among others. Appendix A gives an overview of the ageing models used in this work.

When the AGN is active, it accelerates charged particles, resulting in their having a power law energy distribution. The particles lose energy mainly by radiating synchrotron radiation and through IC scattering off of CMB photons. The radio spectrum has a spectral index $ \alpha_{0} $ at low frequencies, which depends on the energy distribution of the radiating particles and which is commonly known as the injection (spectral) index. Typical values for the injection spectral index are $ \alpha_{0} \in [-0.6, -0.8] $. At high frequencies, due to the preferential cooling of high energy electrons through synchrotron radiation and IC scattering, the radio spectrum steepens, and a spectral break develops. After the AGN shuts down, the active radio regions stop being replenished with energetic particles, and the spectrum exhibits a steep drop-off at high frequencies. 

\begin{figure*}[!htpb]
\captionsetup[subfigure]{labelformat=empty}
\centering
\subfloat[][]{\includegraphics[width=50mm]{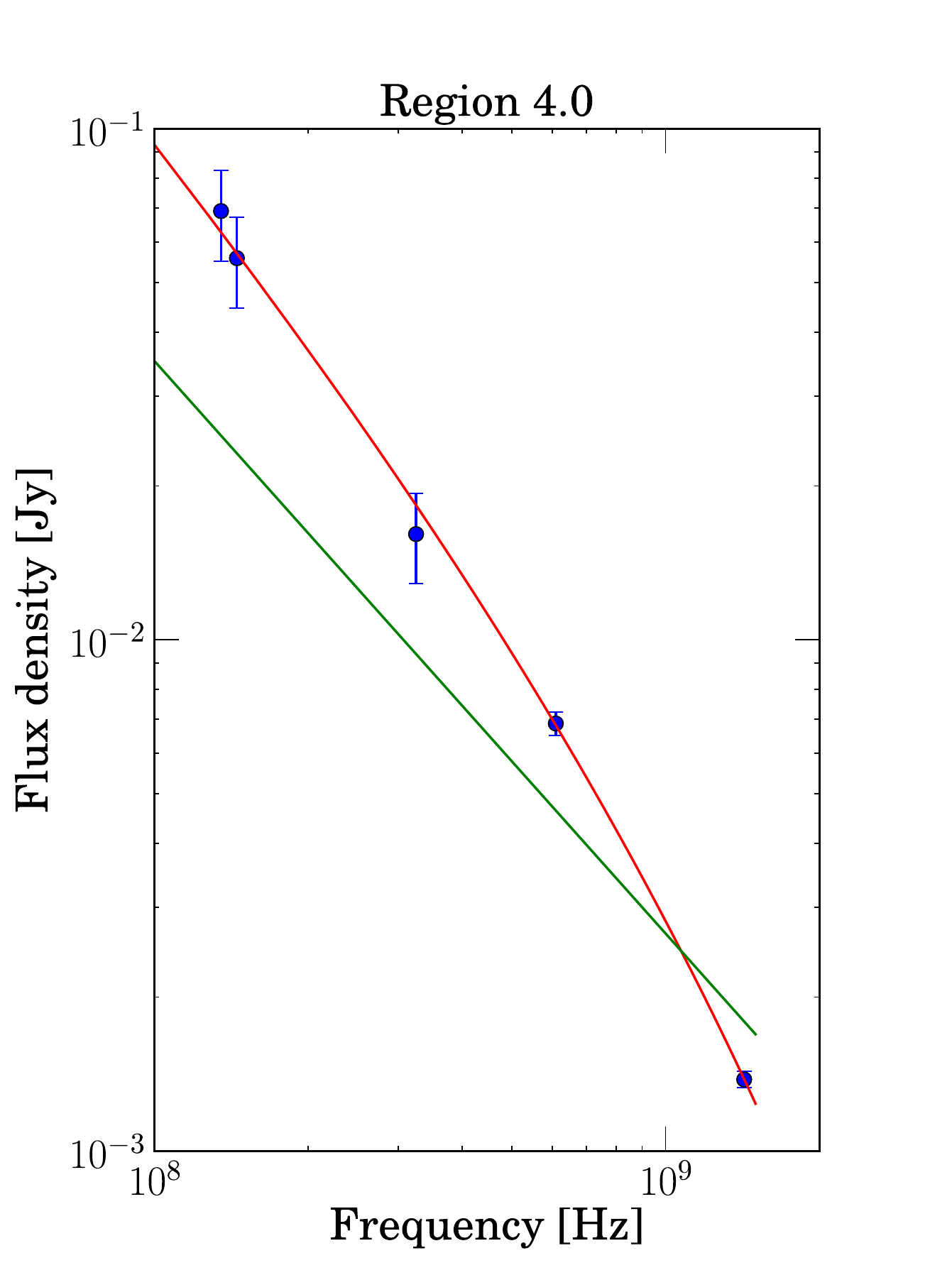}}
\subfloat[][]{\includegraphics[width=90mm]{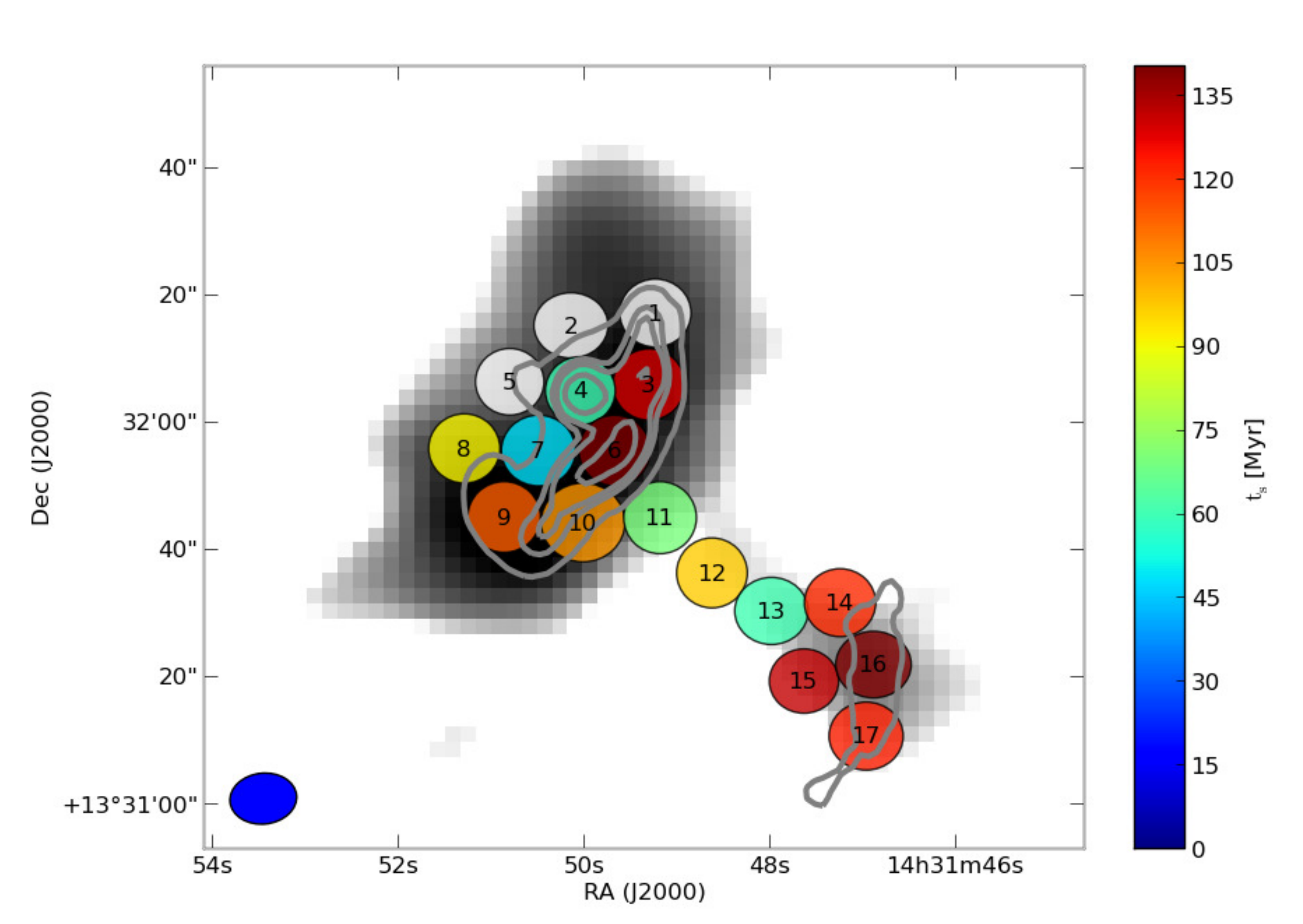}}
\subfloat[][]{\includegraphics[width=41mm]{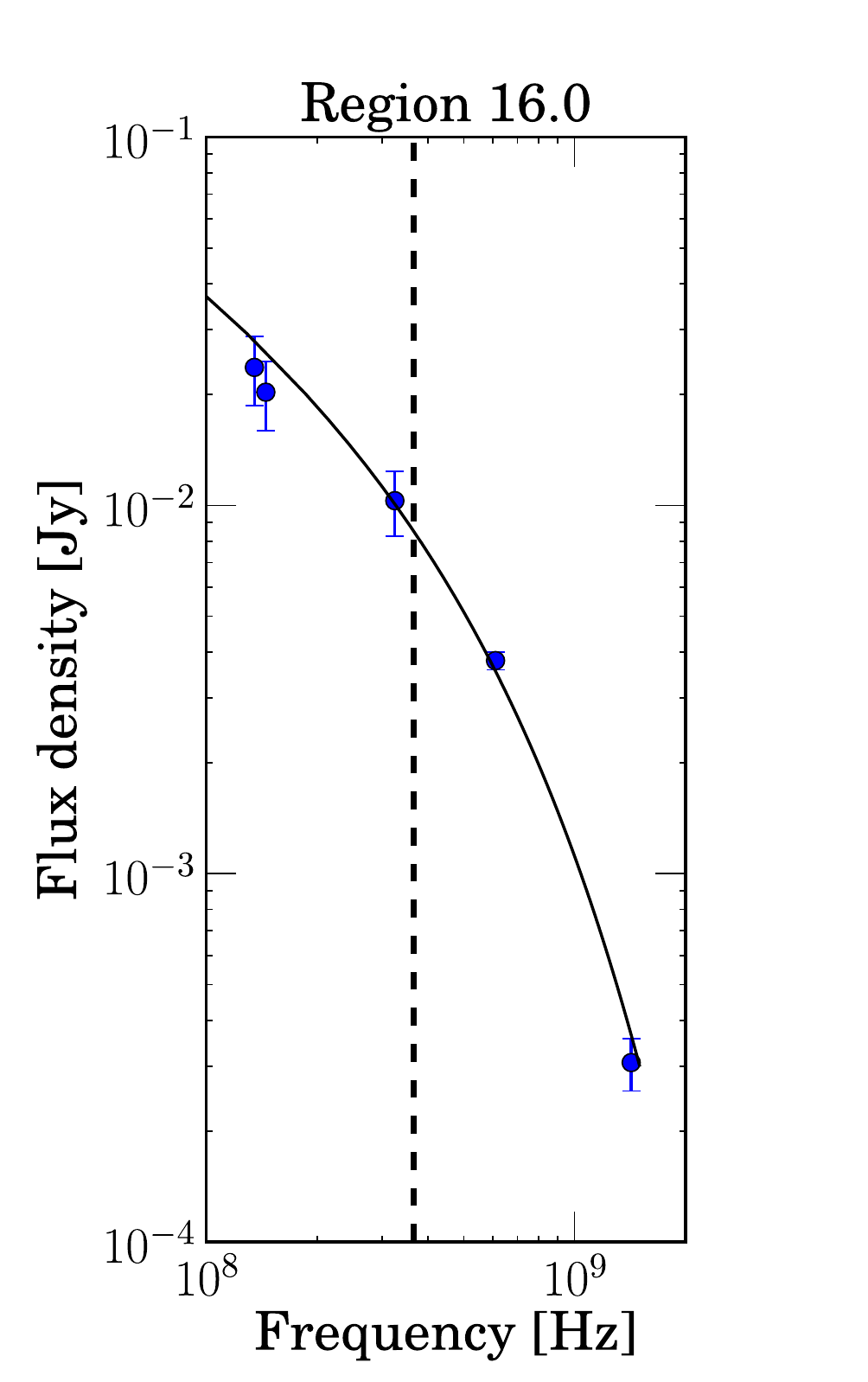}}
\caption{Derived total ages ($ t_{\mathrm{s}} = t_{\mathrm{on}} + t_{\mathrm{off}}  $) for different areas using ageing model (JP and KGJP) fits for the data. The areas are marked with coloured ellipses overlaid on a LOFAR greyscale map of the target. Overlaid are also VLA 1425 MHz high resolution contours in grey (same as in Figure \ref{c4:tgt_regions}) indicating the position of the radio core in Area 4. Areas for which all the model fits were rejected are indicated in white. Model fits for Areas 4 and 16 are shown in the left and right panels, respectively. The green line in the plot for Area 4 represents the (rejected) CI model fit. The break frequency for Area 16 ($ \sim 380 $ MHz) is shown with a dashed black vertical line. Detailed fit results are shown in Table \ref{c4:table:3}.}
\label{c4:reg_ages}
\end{figure*}

To get an estimate on the ages of the different regions, we fitted synchrotron ageing models to the data. The fitting was performed using the Kapteyn package \citep{RefWorks:245} utilizing a Python based code that implements the models (see Appendix A for details). We used models with a Jaffe-Perola energy-loss term, assuming instantaneous particle injection and continuous particle injection followed by ageing (JP and KGJP, respectively) to estimate the ages. We also tried a continuous-injection-only (CI) model for the source area containing the radio core (Area 4) and found it to be inconsistent with the data (Figure \ref{c4:reg_ages}).

Determining the radiative ages requires knowledge about the magnetic field. We calculated it according to \cite{RefWorks:14} using the assumption of equipartition between the energy contained in the field and relativistic particles, using values of $ 10 $ MHz and $ 10 $ GHz for the cutoff frequencies and an electron-to-proton ratio equal to unity. The calculation was done for Regions A and B separately, taking the path length through the regions to equal the average of their major and minor axes as projected on the sky. We set the spectral index to $ \alpha_{144}^{1425} \, = \, -1.9 $ for both regions for the purposes of this calculation \citep[estimated from our spectral index map and the ones given in][]{RefWorks:145}. Using the average value of the derived magnetic field for both regions, we find that $ B \, = \, 4.4 \, \mu $G.

This equipartition value for the magnetic field needs to be corrected, taking a cutoff in the particle energy into account instead of a cutoff in the frequency of the radiation \citep{RefWorks:272}. We adopted the method proposed by \cite{RefWorks:279} \citep[see also][]{RefWorks:273}. The ratio of energy between protons and electrons is assumed be unity. Furthermore, we used spectral index values of $ -1 $ and $ -1.8 $ over the entire energy band, as well as two different values for the low energy cutoff, expressed via the electron Lorentz factor $ \gamma $. The derived magnetic field values are given in Table \ref{c4:table:mag}. Different parameter assumptions change the magnetic field by a factor of 3. We adopt a magnetic field value of $ 4 \, \mu $G in our further analysis, which agrees with what is found for similar sources \citep{RefWorks:34}.

\begin{table}[!htpb]
\noindent \caption{\small Equipartition magnetic field $ B_{\mathrm{eq}} $ and corresponding corrected values $ B_{\mathrm{eq}}^{\mathrm{cor}} $.}
\label{c4:table:mag}
\begin{tabular}{p{2cm} p{2cm} p{2cm} p{1cm}}
\hline\hline\\
$ \alpha $ & $ \gamma_{\mathrm{min}} $ & $ B_{\mathrm{eq}} $[$\mu$G] & $ B_{\mathrm{eq}}^{\mathrm{cor}} $[$\mu$G] \\
\hline\\
-1\tablefootmark{a} & 100 & 2.18 & 3.86 \\ 
-1\tablefootmark{a} & 700 & 2.18 & 2.37 \\ 
-1.8 & 100 & 4.40 & 11.26 \\ 
-1.8 & 700 & 4.40 & 3.925 \\ 
\hline
\end{tabular}
\tablefoot{
The IC equivalent magnetic field for the redshift of J1431.8+1331 is $ B_{\mathrm{IC}} = 3.57 \, \mu $G. \\
The minimum magnetic field (resulting in maximum radiative ages) is $ B_{\mathrm{min}} = 2.52 \, \mu $G. \\
The correction to the derived equipartition magnetic field value is performed according to \cite{RefWorks:279} and \cite{RefWorks:273}. \\
\tablefoottext{a}{Injection spectral index.}
}
\end{table}

\begin{table}[!htpb]
\noindent \caption{\small Best fit $ t_{\mathrm{on}} $, $ t_{\mathrm{off}} $ and $ \alpha_{0} $ values for different source regions.}
\label{c4:table:3}
\begin{tabular}{p{0.8cm} p{1.9cm} p{1.9cm} p{1.8cm} p{0.6cm}}
\hline\hline\\
$ N^{o}_{=} $ & $ t_{\mathrm{on}} $ [Myr] & $ t_{\mathrm{off}} $ [Myr] & $ \alpha_{0} $ & $ \chi^{2} $ \\
\hline\\
\multicolumn{5}{l}{Region A}\\
\hline\\
3 & $ 20.4 \pm 0.0 $ & $ 108.4 \pm 2.8 $ & $ -0.88^{+0.28}_{-0.38} $ & 0.17 \\
4 & - & $ 60.0 \pm 3.5 $ & $ -1.12 \pm 0.38 $ & 0.72 \\
6 & $ 22.9 \pm 6.3 $ & $ 117.4 \pm 0.0 $ & $ -0.74^{+0.14}_{-0.33} $ & 0.47 \\
7 & - & $ 48.6 \pm 5.0 $ & $ -1.50^{+0.47} $ & 0.23 \\
8 & - & $ 91.3 \pm 9.3 $ & $ -1.50^{+0.47} $ & 1.72 \\
9 & - & $ 113.2 \pm 3.5 $ & $ -1.03^{+0.38}_{-0.33} $ & 0.01 \\
10 & - & $ 104.7 \pm 2.7 $ & $ -0.93^{+0.33}_{-0.28} $ & 0.01 \\
\hline\\
\multicolumn{5}{l}{Bridge}\\
\hline\\
11 & - & $ 69.8 \pm 10.9 $ & $ -1.50^{+0.43} $ & 2.72 \\
12 & - & $ 97.2 \pm 0.0 $ & $ -1.50^{+0.66} $ & 1.91 \\
13 & - & $ 60.6 \pm 13.9 $ & $ -1.50^{+0.81} $ & 1.25 \\
\hline\\
\multicolumn{5}{l}{Region B}\\
\hline\\
14 & - & $ 121.4 \pm 5.9 $ & $ -0.60_{-0.29} $ & 1.49 \\
15 & - & $ 131.6 \pm 13.5 $ & $ -0.60_{-0.66} $ & 0.51 \\
16 & $ 7.8 \pm 7.0 $ & $ 132.4 \pm 2.8 $ & $ -0.6 $ & 4.98 \\
17 & - & $ 123.1 \pm 7.5 $ & $ -0.6_{-0.14} $ & 2.34 \\
\hline
\end{tabular}
\tablefoot{
The region age is: $ t_{\mathrm{s}} = t_{\mathrm{on}} + t_{\mathrm{off}} $. For regions where $ t_{\mathrm{on}} $ is not listed, the ageing only (JP) model provided the best fit, in which case $ t_{\mathrm{s}} \, = t_{\mathrm{off}} $\\
We test the acceptance of the hypothesis that the data is consistent with the model using a significance level with a probability of 5\%. The $ \chi^{2} $ threshold for the 95\% confidence interval \citep{RefWorks:197} for two degrees of freedom is $ \chi^{2}_{95\%} = 5.99 $.}
\end{table}

We fitted the JP and KGJP models to our data for each area of the source using the following procedure. The free parameters of the model fit were the active and switched-off times, as well as a flux scaling factor, while we kept the magnetic field and injection spectral index fixed. To understand the injection spectral index dependence, we took $ 20 $ different values for the injection spectral index evenly distributed between $ -0.6 $ and $ -1.5 $ and performed the model fitting for each source area separately using different injection spectral index values.

We find that the best fit to the data is obtained for the models that use the JP assumption for the energy loss. Most of the source areas are best fitted with a model that assumes an infinitesimally short duration particle acceleration phase and ageing (JP), while some are best fitted with a model that assumes a period of continuous injection and then ageing (KGJP). Our best-fit parameter values are given in Table \ref{c4:table:3}, and the derived ages for the various source areas are presented in Figure \ref{c4:reg_ages}. The confidence intervals for the injection spectral index parameter were obtained by considering all of the injection spectral index values for which the model was accepted (irrespective of the $ \chi^{2} $ value of the fit). The adopted best-fit value of the injection spectral index was the value used in the model fit with the lowest $ \chi^{2} $ value. The ages and their corresponding confidence intervals were the ones obtained by the model fit using the best-fit injection spectral index value.

For Regions 7, 8, 11, 12, and 13, the best-fit injection index was the steepest in the range, and we could only place an upper limit on it, while the inverse was true for Regions 14, 15, and 17. For Region 16, the only model fit that was not rejected was the one with an injection index of $ \alpha_{0} = -0.6 $. With all this in mind, judging from the regions for which we have constrained the injection index, we deduce that the youngest part of the source is located in Region A, in the area around the faint radio core (Region 4), while the oldest source region is Region B. The derived ages range from 60 to 130 Myr. The models indicate that for Region B the injection spectral index has a value of $ \alpha_{0} \, = \, -0.6 $, while for Region A, the best fit models are those with the injection index of around $ \alpha_{0} \, = \, -1 $. We discuss the implications of these findings further in Section \ref{c4:disc}.

\subsubsection{Colour-colour representation and shift diagrams}
\label{c4:colour-shift}

The peculiar morphology of the source suggests that complex processes are shaping the emission regions and influencing their energy balance. We investigate whether the energy loss is predominantly due to synchrotron ageing and IC losses.

To do this and to gain more insight into the plasma properties across the source, we employed the colour-colour representation described by \cite{RefWorks:189}. Such plots provide a simple overview of the measurements compared to different ageing models and allow for easier determination of which theoretical models describe the measurements best. Also, the outlined shapes in this representation are conserved with respect to changes in the magnetic field and adiabatic compression or expansion.

\begin{figure*}[!htpb]
\captionsetup[subfigure]{labelformat=empty}
\centering
\subfloat[][]{\includegraphics[width=0.52\textwidth]{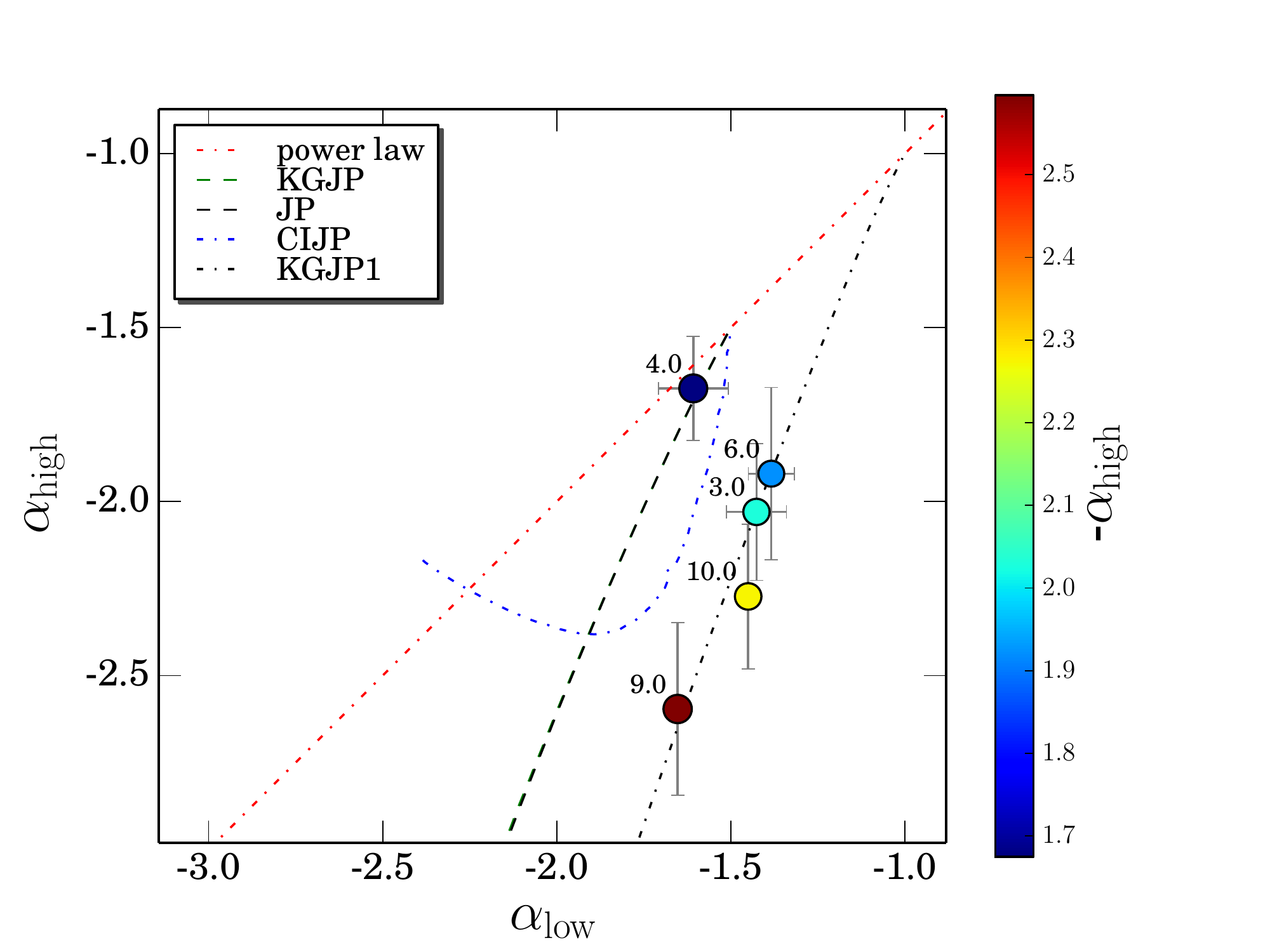}}
\subfloat[][]{\includegraphics[width=0.52\textwidth]{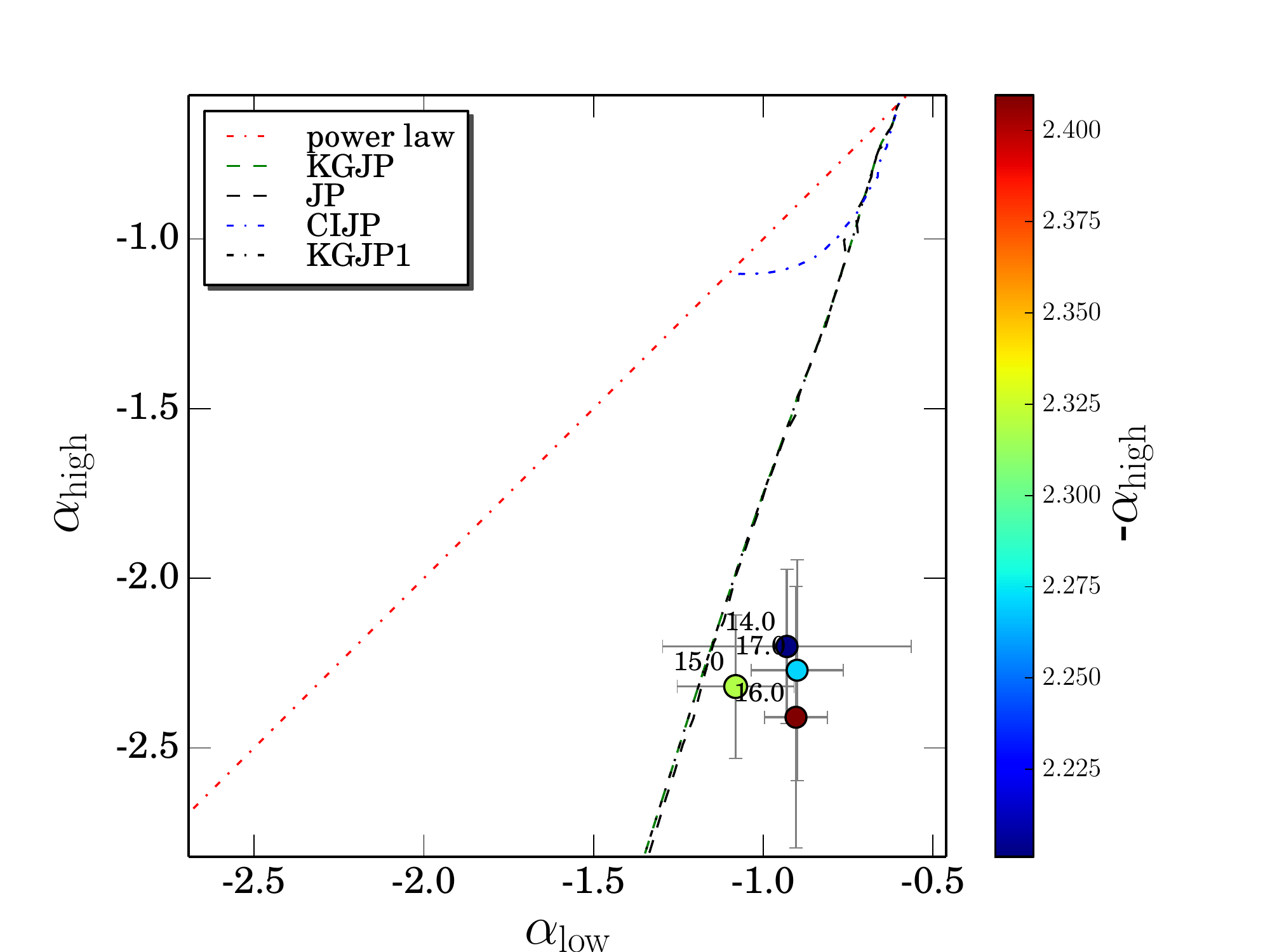}}
\caption{Colour-colour plots for different regions across the source. The dash-dotted red line indicates the locus of points for a power-law spectrum. \textbf{Left:} The dashed green line and the solid black line (overlapping) represent the locus of points taken by KGJP and JP models, respectively, with an injection index of $ \alpha_{0} \, = \, -1.5 $. The dash-dotted blue line represents a CIJP model with the same injection index. The dash-dotted black line represents a KGJP model with injection index of $ \alpha_{0} \, = \, -1 $. \textbf{Right:} The dashed green line and the dashed black line (overlapping) represent the locus of points taken by KGJP and JP models, respectively, with injection index of $ \alpha_{0} \, = \, -0.6 $. The dash-dotted blue line represents a CIJP model with the same injection index. Marker sizes are proportional to the low-frequency spectral indices for a given region and marker colours are proportional to the high-frequency spectral indices. Numbers label the regions.}
\label{c4:colour-colour}
\end{figure*}

We have constructed two spectral indices from our data, low and high. The low spectral index was obtained by fitting a first-order polynomial through the $ 135 $, $ 145 $, and $ 325 $ MHz data points for each region, and the high frequency spectral index was obtained by a first-order polynomial fit through the $ 325 $, $ 610 $, and $ 1425 $ MHz data points. The resulting colour-colour plots for a representative set of source regions are shown in Figure \ref{c4:colour-colour}. This procedure is similar to the spectral curvature derivation described previously, with the curvature being the difference between the low and high spectral indices. Plotting these two indices against each other is another way of distinguishing source regions with different properties that can give additional insight into the physical processes at work in the emission regions.

In the colour-colour plot for source Region A we have taken only the measurement areas into account for which we have a good constraint on the injection index. We did not perform such a pre-selection for the colour-colour plot for Region B, since in that case we have a more limited number of measurement areas.

We can see that Area $ 4 $ is close to a power law, but that it is best described by KGJP or JP models having a steep injection index (determined by the intersection point of the ageing models and the power law line on the diagram) of $ \alpha_{0} = -1.5 $. Measurement areas $ 3 $, $ 6 $, $ 9 $, and $ 10 $ are best described by a KGJP model with an injection index of $ \alpha_{0} = -1 $. The areas in the source Region B show larger scatter and are not represented well with a single ageing model; a KGJP or JP model with injection index of $ \alpha_{0} < -0.6 $ is indicated by the measurements and supported by the derived injection indices and ages in Section \ref{c4:ages}.

Following the reasoning of \cite{RefWorks:190}, \cite{RefWorks:191, RefWorks:192}, and \cite{RefWorks:193}, we performed a spectral shift analysis on Regions $ 3 $, $ 6 $, $ 9 $, and $ 10 $ to gain further insight into our target and test the assumptions we made during the spectral analysis. We have chosen these regions since they outline a single curve in the colour-colour space that is consistent with a particular ageing model.

\begin{figure*}[!htpb]
\centering
\subfloat[][The slope of the fit is $ -1.6 $.]{\includegraphics[width=0.5\textwidth]{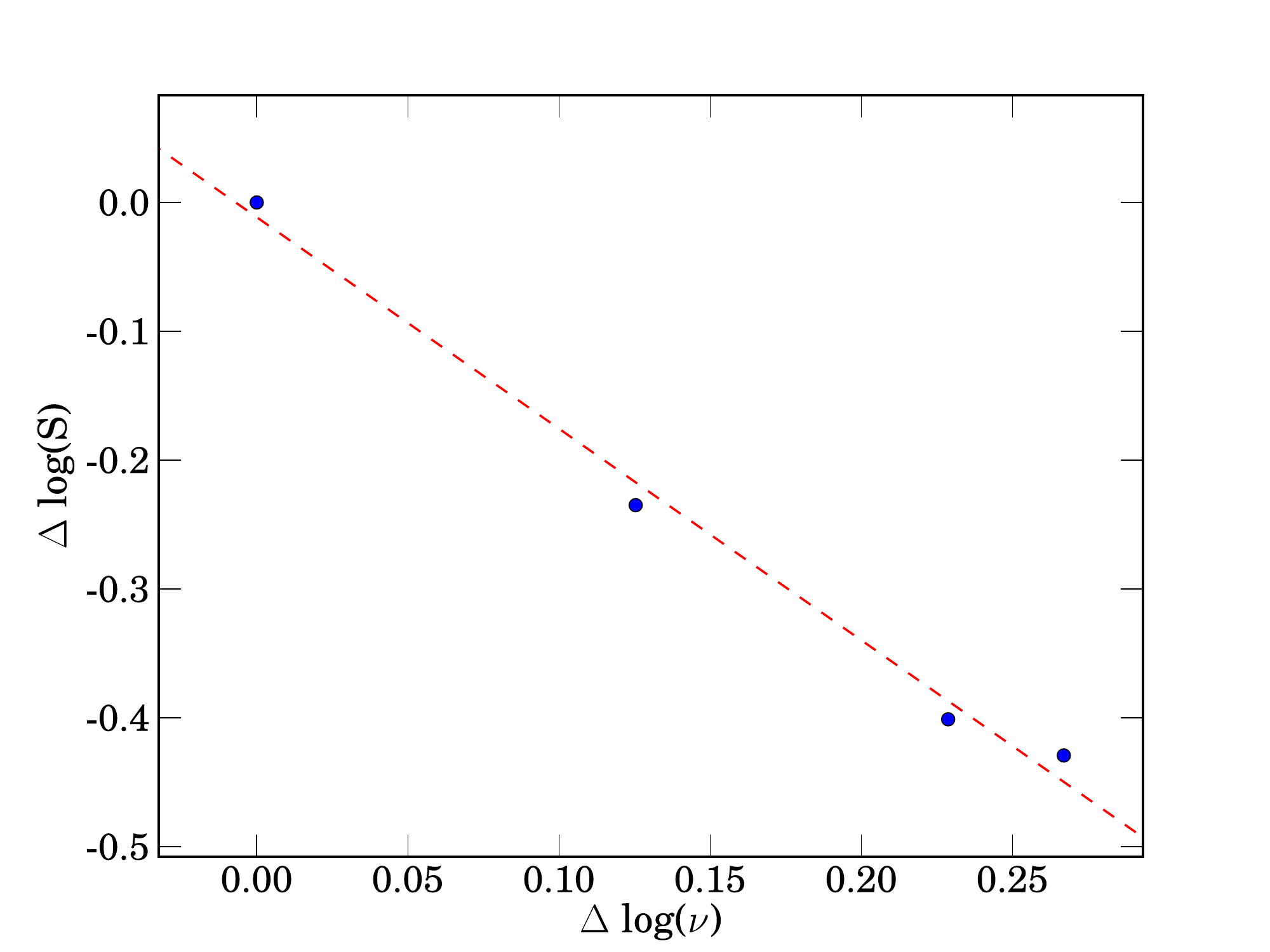}
\label{c4:shifts:subfig1}}
\subfloat[][]{\includegraphics[width=0.46\textwidth]{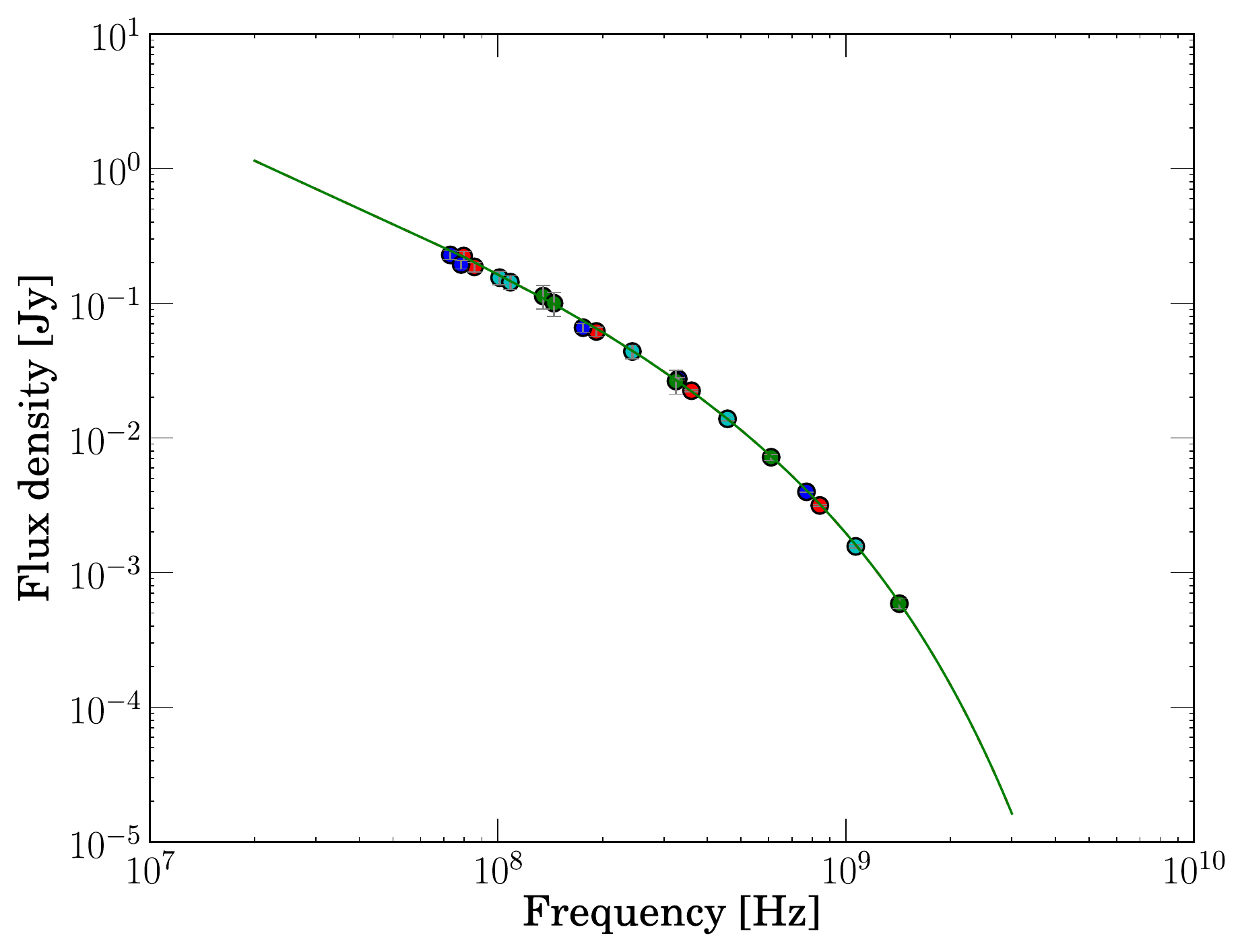}
\label{c4:shifts:subfig2}}
\caption{Spectral shifts in $ \log(\nu) - \log(S) $ for Regions 3, 6, 9, and 10 (\ref{c4:shifts:subfig1}). The dashed red line shows the best linear fit to the data points. Shifted data points giving the global spectrum are given in panel \ref{c4:shifts:subfig2}. Different colours are used to indicate the data points belonging to different regions. The best JP and KGJP model fits are plotted with dashed-dot red and green lines, respectively (overlapping, since there is essentially no difference between the models for these regions).}
\label{c4:shifts}
\end{figure*}

The idea behind the shift technique is as follows. Assuming that the energy loss of the particles is predominantly through synchrotron and IC mechanisms and that the magnetic field does not vary significantly across the source, then the spectra for all of the regions are self-similar, the only difference between them being the position of the break frequency. Moreover, all of the spectra should align by shifting them in the frequency-flux density plane if the dominant energy loss mechanism is synchrotron radiation and IC scattering and if our line of sight samples similar source regions. We chose Region $ 9 $ as reference (it has the best model fit) and have shifted the spectra for Regions $ 3 $, $ 6 $, and $ 10 $ so that their break frequencies matched that of Region $ 9 $. The break frequencies for the KGJP model were determined using \citep{RefWorks:34}:

\begin{equation}
\label{c4:ageq}
\nu_{\mathrm{b}}^{\mathrm{low}} = \left(\frac{1590}{(B^{2} + B_{\mathrm{CMB}}^{2})t_{\mathrm{s}}}\right)^{2} \cdot \frac{B}{(1 + z)}
\end{equation}

\noindent and

\begin{equation}
\nu_{\mathrm{b}}^{\mathrm{high}} = \nu_{\mathrm{b}}^{\mathrm{low}}\left( \frac{t_{\mathrm{s}}}{t_{\mathrm{off}}} \right)^{2}
\end{equation}

\noindent where $ t_{s} \, = \, t_{on} \, + \, t_{off} $. For the JP model, we set $ t_{on} \, = \, 0 $.
The procedure has produced the shift diagram for these regions, shown in the left-handed panel of Figure \ref{c4:shifts}. It shows by how much the spectrum of each region was shifted in the $ \log(\nu) - \log(S) $ plane. The data points for all of the regions after the shift, producing the global spectrum, are shown in the right-handed panel of Figure \ref{c4:shifts}. We also fitted JP and KGJP models (Appendix A) through all of the shifted data points to show that indeed the shifted spectra are self-similar and taken together produce a spectrum that is consistent with radiative ageing and IC scattering being the dominant energy loss mechanism of the electrons.

The shifts made to the individual spectra to line them up can give us information about the physical parameters of the source \citep{RefWorks:190, RefWorks:193}. Specifically, shifts in the $ log(\nu) $ axis relate to $ \gamma^{2}B $, while shifts along the $ log(S) $ axis depend on $ N_{\mathrm{tot}}B $, where $ N_{\mathrm{tot}} $ is the total number of energetic particles in the source volume for a given beam size.

We can fit a straight line to the shifts in the $ log(\nu) - log(S) $ plane (Figure \ref{c4:shifts}) for the measurement areas belonging to source component A. This means that the spectra of individual regions are self-similar, meaning that there are no significant variations in either particle energy or magnetic field value from one area to another, and that the spectral shapes are the result of synchrotron ageing and IC losses. Moreover, the slope of the fit indicates the injection index.

That the shifted data are fitted so well using radiative ageing models shows that, indeed, synchrotron ageing and IC scattering are the dominant energy loss mechanisms. The values we get for the injection index and that they match what we had used for $ \alpha_{0} $ previously also serve to strengthen the inner consistency of our analysis.

\subsubsection{The steepness of the injection index}
\label{c4:steepinj}

The derived results on the ages of source Region A are consistent with it being a relic from a past AGN activity episode; this claim is mainly supported by the derived age of Area $ 4 $ compared to its surroundings and its identification with the faint radio core. There is consistency throughout our model fits: for Region A, the models with steeper injection index ($ \alpha_{0} \in [-0.9, -1.5] $) than what is usually accepted as standard value ($ \alpha_{0} \in [-0.6, -0.8] $) are the ones that best fit the data. We discuss the implications of the steep injection index further in Section \ref{c4:disc}.

Interestingly, \cite{RefWorks:251} find larger-than-expected injection indices for their sample of FRII radio galaxies.

\subsubsection{The integrated flux density spectrum}
\label{c4:intspec}

The integrated spectrum is slightly curved and steep. To see how spectral ageing in different regions of the source can affect the integrated spectrum, we fitted a KGJP model to the integrated flux density measurements, and compared the results with the region fits done previously.

\begin{figure}[!htpb]
  \centering
  \includegraphics[width=75mm]{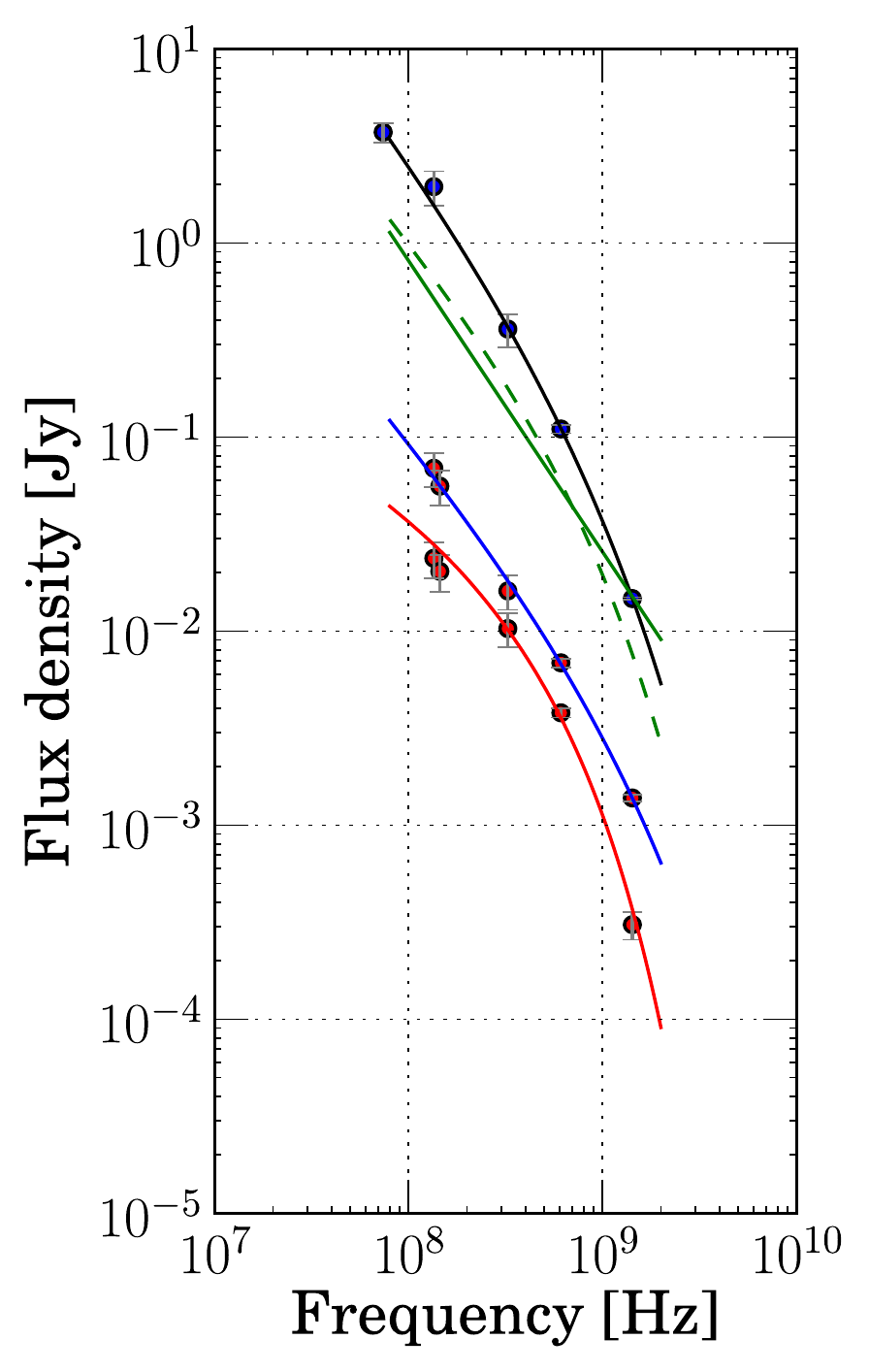}
  \caption{JP and KGJP model fits for Regions 4 (blue) and 16 (red), along with the corresponding data points (LOFAR HBA, GMRT, and VLA). The sum of the fits for all regions is represented by the dashed green line. The integrated flux density data points (VLSS, LOFAR HBA, GMRT, and VLA) are best fitted by a KGJP model (black line) with an injection index of $ \alpha_{0} \, = \, -1.23 \pm 0.37 $ and $ t_{\mathrm{on}} \, = \, 0.1 $ Myr, $ t_{\mathrm{off}} \, = \, 77.7 $ Myr. The measurement regions do not cover the source completely at lower frequencies, so the sum of the fits does not precisely follow the integrated flux density. A CI model that is rejected by the integrated flux data is represented with a green line.}
\label{c4:int_spec}
\end{figure}

In Figure \ref{c4:int_spec} we show the integrated flux density measurements of VLSS J1431.8+1331 from our LOFAR HBA, GMRT, and VLA data sets along with the VLSS catalogue value. We can see that the integrated flux density follows the sum of the fitted flux densities for the different regions; the offset can be explained by the fact that our measurement regions do not perfectly cover the entire surface of the target. The spectra of different regions have breaks at different frequencies and have different curvatures. When they combine, the effect is that the integrated spectrum has smaller curvature. This demonstrates that we should be careful when interpreting integrated spectral shapes of complex sources. The activity history, coupled with the detailed source morphology and the relative contribution of different source components can influence the shape of the integrated spectrum. We discuss the implications of this further in Section \ref{c4:disc}.

\section{Discussion}
\label{c4:disc}

The spectral index maps we derived using LOFAR data, which extend a factor of two lower in frequency than previous studies of this object, confirm that there is a frequency break in the spectrum of source Region B, while the larger one, Region A, still retains a steep spectral index at lower frequencies.

Our spectral mapping is in good agreement with the findings of \cite{RefWorks:145}. It confirms that the spectral index of Region B derived at lower (LOFAR HBA) frequencies continues to flatten out and reaches injection values. We are observing the particles that still have kept their energy from the last episode of acceleration. The data suggest a spectral break in the spectrum of this region at relatively low frequencies. This conclusion is supported by the spectral curvature map in Figure \ref{c4:spix_curv_maps}.

Our synchrotron ageing models indicate that the youngest radio plasma in the target (around 60 Myr old) is located in the vicinity of the fading AGN core (Area 4 in Figure \ref{c4:reg_ages}). This is what would be expected if the AGN was diminishing in activity over time for a FRI-like radio source morphology. The rest of Region A (apart from Areas 4, 7, and 8) is older (around 100 Myr). The oldest part of the source is Region B, with ages ranging from 120 Myr to 130 Myr. What we are seeing, then, could be a signature of an AGN hosted by the BCG of the cluster that was active around 120 Myr ago when it produced source Region B and the outer part of Region A. Then, the AGN switched to very low levels of activity around 60 Myr ago.

We can use morphological and kinematic arguments as an independent check of the timescales involved. The relative speed of maxBCG~J217.95869+13.53470 with respect to the group of galaxies to the north-east is around 3200 \kms\ \citep{RefWorks:86}. It is suspected of being a separate smaller cluster; one of its galaxies hosts a spectrally confirmed, currently active AGN. We assume that this value is equally distributed between the two clusters and that the position of the BCG of maxBCG~J217.95869+13.53470 at the beginning of the outburst was at the half-way point between Regions A and B. Then, the activity began to decrease as the BCG followed the cluster motion from south-west to north-east. The motion of the Intra Cluster Medium (ICM) halo gas (centred on the cluster core) has bent the radio emission, giving it its current appearance. In this scenario, the distance from the assumed position where the AGN in the host galaxy started its previous outburst of activity to the present position of the galaxy is 104 \kpc. Moving at the previously mentioned speed, it would take 32 Myr for the AGN host galaxy to arrive to its present position, which is around four times shorter than the estimated age of Region B. Large uncertainties in the cluster velocity estimates propagate into the kinematic timescale, and we should keep this in mind.

What does the difference in the observed injection spectral indices between the two source regions mean? If we assume that Region A was produced as a result of an AGN activity, the steeper injection spectral indices we infer for its plasma broadly agree with what was observed for several radio galaxies by \cite{RefWorks:251} (The radio galaxies in their sample are FRIIs, while J1431.8+1331 was probably an FRI radio galaxy.) Region B would be plasma released in a previous episode of AGN activity that has faded, losing its energy through radiation and expansion. \cite{RefWorks:86} estimate that it would take Region B around 300 Myr to get to its present position, rising as a buoyant bubble. It could have been compressed by a merger shock produced by the interaction of the two galaxy clusters (the BCG of one that hosts J1431.8+1331). The shock compression would cause the plasma bubble to gain energy, resulting in a boost in its radio emission. 

The compression of the plasma in Region B would lead to the formation of a sheath-like radio emission region, which as time passes, would evolve into a filamentary structure \citep{RefWorks:146}. When looking at the radio morphology of Region B, this is a possible scenario if we assume that we see it nearly edge on.

In relation to the compression shock scenario, another explanation of the source spectral properties presents itself. Its plausibility increases if we look at the injection spectral index values for the different measurement regions given in Table \ref{c4:table:3}. The regions for which we only have upper limits to the injection index and which have the steepest injection index are situated in Region A and in between Regions A and B. This may point to the possibility that what we are actually observing is the spectrum above the break frequency for Region A, meaning that the ages we derive for it are lower limits to the true age. In this context, Region B can be even older, and the shock compression could have not only revived its radio emission but also shifted its break frequency to the higher values that we observe \citep{RefWorks:274}.

We can estimate the maximum electron ages using Eq. (\ref{c4:ageq}) by taking the minimum value of the magnetic field of $ B_{\mathrm{min}} = \frac{B_{\mathrm{CMB}}}{\sqrt{3}} = 2.52 \, \mu $G. Assuming a value for the break frequency of 60 MHz, we get maximum ages of around $ 350 $ Myr, which make the scenario mentioned above plausible. Under the same assumptions, but using a magnetic field value of $ 10 \, \mu $G, we find an age of $ 160 $ Myr, which is more than a factor of two smaller. As can be expected, with everything else being equal, uncertainties in the magnetic field determination (Table \ref{c4:table:mag}) can affect the derived ages.

As we have shown in Section \ref{c4:disc}, the integrated flux density spectrum is steep and curved. This is caused by superposition of different source regions that have spectral breaks at different frequencies. It is similar to the spectra of the cluster sources observed by \cite{RefWorks:34}. The integrated source spectrum contains information about the detailed spectral properties, depending on which source region is dominating the radio emission. In the case of J1431.8+1331, the steep-spectrum, larger region is the dominant one, so the overall spectrum is very steep at higher frequencies, showing characteristics of a fading radio source spectrum.

\section{Conclusions}
\label{c4:fin}

We have observed the steep-spectrum radio source VLSS~J1431.8+1331 using LOFAR, adding to an already existing data set taken with the GMRT and the VLA by \cite{RefWorks:145}. Using the resulting broad band frequency coverage and high resolution imaging, we were able to study the spectral properties of different regions across the source.

Based on our analysis, we conclude that the AGN in Region A shut down around 60 Myr years ago.

We found that different source regions have different spectral properties and ages. Our results are consistent with the results of \cite{RefWorks:86}, who suggest that source Region B might be a shock compressed plasma bubble.

The integrated flux density spectrum can be used as a classifying tool to identify (unresolved) sources in surveys that are comprised of multiple regions resulting from different stages of AGN activity.

\begin{acknowledgements}

LOFAR, the Low Frequency Array designed and constructed by ASTRON, has facilities in several countries, which are owned by various parties (each with their own funding sources), and which are collectively operated by the International LOFAR Telescope (ILT) foundation under a joint scientific policy.\\
RM gratefully acknowledges support from the European Research Council under the European Union's Seventh Framework Programme (FP/2007-2013) / ERC Advanced Grant RADIOLIFE-320745.\\
This research made use of the NASA/IPAC Extragalactic Database (NED), which is operated by the Jet Propulsion Laboratory, California Institute of Technology, under contract with the National Aeronautics and Space Administration.\\
This research made use of APLpy, an open-source plotting package for Python hosted at http://aplpy.github.com\\
RJvW acknowledges the support by NASA through the Einstein Postdoctoral grant number PF2-130104 awarded by the Chandra X-ray Center, which is operated by the Smithsonian Astrophysical Observatory for NASA under contract NAS8-03060.\\
We thank the anonymous referee for the constructive comments that have improved the manuscript.

\end{acknowledgements}

\section*{Appendix A: Synchrotron emission models}
\label{c4:appb}

The fundamentals of the formalism describing the synchrotron emission mechanisms were given by \cite{RefWorks:195} and later expanded by \cite{RefWorks:126} in his seminal paper discussing various particle injection and energy loss processes, including synchrotron radiation. \cite{RefWorks:187} and \cite{RefWorks:125} expanded their analysis further.

Assuming that the radiating particles have energies described by a power law with a spectral index $ \gamma $: $ E^{-\gamma} $, they produce a radio spectrum that is also described by a power law with a spectral index $ \alpha = -\frac{\gamma - 1}{2} $. As time passes, these particles lose energy through synchrotron radiation and inverse Compton (IC) scattering off the CMB photons. These losses can be represented by an energy loss term, depending on the model used. The Kardashev - Pacholczyk (KP) model assumes that the pitch angle of the radiating particles (the angle between the instantaneous velocity vector of the particle and the magnetic field lines) remains constant over time, giving a loss term:

\[ b = c_{2} B^{2}\left[sin^{2} \theta + \left(\frac{B_{\mathrm{IC}}}{B}\right)^{2}\right] \]

\noindent where $ \theta $ is the pitch angle, $ B_{\mathrm{IC}} = \sqrt{\frac{2}{3}}B_{\mathrm{CMB}} $ is the effective IC magnetic field, and $ B_{\mathrm{CMB}} = 3.25(1 + z)^{2} $ is the equivalent CMB magnetic field \citep{RefWorks:196}. Here, $ c_{2} = 2.37\cdot10^{-3} $ (CGS units) is a constant defined in Pacholczyk (1971), and $ B $ is the magnetic field in the radiating region. In the Jaffe - Perola (JP) model, it is assumed that the pitch angle is isotropized on timescales much shorter than the radiation timescale, and the loss term becomes

\[ b = c_{2} B^{2}\left[\frac{2}{3} + \left(\frac{B_{\mathrm{IC}}}{B} \right)^{2}\right] \]

For the particle energy we have $ E = c_{1} \sqrt{\nu/x} $ where $ c_{1} = 4\cdot10^{-10}/\sqrt{B} $ (CGS units) is a constant, $ \nu $ the observing frequency, and $ x = \nu/\nu_{\mathrm{b}} $ is the scaled frequency, depending on $ \nu_{\mathrm{b}} $, the frequency of the observed spectral break.

We need to find the particle distribution function, which depends on the energy of the particles. There are several separate cases we should consider. In the simplest case, there is a particle injection episode which lasts for an infinitesimally short time (the injection happens at $ t=0 $), and afterwards the particles radiate away their energy \citep{RefWorks:126}:

\[ N(t_{\mathrm{OFF}},b,\gamma,E) = \left\{ \begin{array}{ccc} E^{-\gamma}(1 - bEt_{\mathrm{OFF}})^{\gamma - 2} & \mbox{for} & E < \frac{1}{bt_{\mathrm{OFF}}} \\ \\ 0 & \mbox{for} & E \geqslant \frac{1}{bt_{\mathrm{OFF}}} \end{array} \right. \]

\noindent We label this model JP or KP depending on the treatment of the loss term, $ b $, we adopt. Here, $ t_{\mathrm{OFF}} $ is the time elapsed since the injection or, in this case, the source age $ t_{s} = t_{\mathrm{OFF}} $.
When there is AGN activity (or shock particle acceleration) that is ongoing, then the distribution function becomes

\[ N(t_{\mathrm{ON}},b,\gamma,E) = \left\{ \begin{array}{ccc} E^{-\gamma}t_{\mathrm{ON}} & \mbox{for} & E < \frac{1}{bt_{\mathrm{ON}}} \\ \frac{E^{-(\gamma + 1)}}{b(\gamma - 1)} & \mbox{for} & E \geqslant \frac{1}{bt_{\mathrm{ON}}} \end{array} \right. \]

This is a continuous injection model, and we label it CIJP or CIKP. In this case, the source age $ t_{s} = t_{ON} $.
Finally, if there is a period of injection of energetic particles followed by a cessation of the activity, we have \citep{RefWorks:188, RefWorks:54}

\[ N(t_{\mathrm{OFF}},t_{\mathrm{ON}},b,\gamma,E) = \]
\[ \left\{ \begin{array}{ccc} \frac{E^{-(\gamma + 1)}}{b(\gamma - 1)((1 - bE(t_{\mathrm{OFF}} - t_{\mathrm{ON}}))^{\gamma - 1} - (1 - bEt_{\mathrm{OFF}})^{\gamma - 1})} & \mbox{for} & E < \frac{1}{bt_{\mathrm{ON}}} \\ \frac{E^{-(\gamma + 1)}}{b(\gamma - 1)(1 - bE(t_{\mathrm{OFF}} - t_{\mathrm{ON}}))^{\gamma - 1}} & \mbox{for} & \frac{1}{bt_{\mathrm{OFF}}} \leqslant E \leqslant \frac{1}{b(t_{\mathrm{OFF}} - t_{\mathrm{ON}})} \\ 0 & \mbox{for} & E > \frac{1}{b(t_{\mathrm{OFF}} - t_{\mathrm{ON}})} \end{array} \right. \]

This continuous injection model with a switch off we label KGJP or KGKP \citep{RefWorks:54}. Now, we can calculate the observed flux density as

\[ S(\nu) = S_{0}\sqrt{\nu}\int F(x)x^{-1.5}N(x)dx \]

\noindent for the JP case, and

\[ S(\nu) = S_{0}\sqrt{\nu}\int_{0}^{\pi} \sin^{2}\theta \int F(x)x^{-1.5}N(x,\theta)d\theta dx \]

\noindent in the KP case, where $ S_{0} $ is a flux density scaling factor, and $ F(x) = x\int_{x}^{\infty} K_{5/3}(z)dz $ with $ K_{5/3} $ being the modified Bessel function.

\bibliographystyle{aa.bst}
\bibliography{aa.bib}

\end{document}